%% file: main-pqqs.tex
\newif\iflncs
\lncsfalse       

\iflncs
  \documentclass[runningheads]{llncs}
\else
  \documentclass[sigconf,nonacm]{acmart}
\fi

\iflncs
\else
  \setcopyright{none}
  \acmConference[ACM PQQS '26]
    {ACM Conference on Post-Quantum and Quantum-based Security}
    {November 2--4, 2026}
    {San Jose, CA, USA}
  \acmYear{2026}
  \acmDOI{}
  \acmISBN{}
\fi

\usepackage{booktabs}
\usepackage{tabularx}
\usepackage{array}
\usepackage{multirow}
\usepackage{makecell}
\usepackage{amsmath}
\usepackage{enumitem}
\usepackage{microtype}
\usepackage{xcolor}
\usepackage{placeins}
\iflncs
  \usepackage{amssymb}
  \usepackage[hidelinks]{hyperref}
\fi

\newcolumntype{Y}{>{\raggedright\arraybackslash}X}
\newcolumntype{C}{>{\centering\arraybackslash}X}

\iflncs
  \spnewtheorem{attack}{Attack}[section]{\bfseries}{\itshape}
\else
  \newtheorem{property}{Property}
  \newtheorem{attack}{Attack}[section]
  \theoremstyle{remark}
  \newtheorem{remark}{Remark}
\fi
\newcommand{\pparagraph}[1]{\paragraph*{#1.}}
\newcommand{\PPT}{\mathsf{PPT}}
\newcommand{\QPT}{\mathsf{QPT}}
\newcommand{\adv}{\mathcal{A}}
\newcommand{\pk}{\mathsf{pk}}
\newcommand{\sk}{\mathsf{sk}}
\newcommand{\bset}{{0,1}}
\newcommand{\hashf}{H}
\newcommand{\KeyGen}{\mathsf{KeyGen}}
\newcommand{\Eval}{\mathsf{Eval}}
\newcommand{\Sign}{\mathsf{Sign}}
\newcommand{\Verify}{\mathsf{Verify}}
\newcommand{\Setup}{\mathsf{Setup}}
\newcommand{\Prove}{\mathsf{Prove}}
\newcommand{\Com}{\mathsf{Com}}
\newcommand{\Open}{\mathsf{Open}}
\newcommand{\Encaps}{\mathsf{Encaps}}
\newcommand{\Decaps}{\mathsf{Decaps}}
\newcommand{\MLDSA}{\textsf{ML-DSA}}
\newcommand{\SLHDSA}{\textsf{SLH-DSA}}
\newcommand{\MLKEM}{\textsf{ML-KEM}}
\newcommand{\Dilithium}{\textsf{Dilithium}}
\newcommand{\Falcon}{\textsf{Falcon}}
\newcommand{\SPHINCS}{\textsf{SPHINCS+}}

\title{SoK: Stablecoins in the Quantum Era}

\iflncs

  \titlerunning{SoK: Stablecoins in the Quantum Era}
  \authorrunning{P. Chatzigiannis et al.}

  \author{
    Panagiotis Chatzigiannis\inst{1}
    \and
    Duc V. Le\inst{2}
    \and
    Navid Alamati\inst{1}
    \and
    Suvradip Chakraborty\inst{1}
  }

  \institute{
    Visa Research, USA\\
    \email{\{pchatzig,nalamati,suvchakr\}@visa.com}
    \and
    Circle Research, USA\\
    \email{levduc112@gmail.com}
  }

\else

  \author{Panagiotis Chatzigiannis}
  \affiliation{%
    \institution{VISA Research}
    \country{USA}
  }
  \email{pchatzig@visa.com}

  \author{Navid Alamati}
  \affiliation{%
    \institution{VISA Research}
    \country{USA}
  }
  \email{nalamati@visa.com}

  \author{Suvradip Chakraborty}
  \affiliation{%
    \institution{VISA Research}
    \country{USA}
  }
  \email{suvchakr@visa.com}
  
  \author{Duc V. Le}
  \affiliation{%
    \institution{Circle Research}
    \country{USA}
  }
  \email{levduc112@gmail.com}   
\fi

\begin{document}

\iflncs

  \maketitle

  \input{abstract}

  \begin{center}
  \end{center}

\else


\input{abstract}


  \maketitle

\fi

\input{intro}
\input{background}
\input{threats}
\input{taxonomy}
\input{gaps}
\input{conclusion}

\iflncs
  \bibliographystyle{splncs04}
\else
  \bibliographystyle{ACM-Reference-Format}
\fi

\bibliography{references,abbrev3,crypto}

\appendix
\input{appendix}

\end{document}

%% file: abstract.tex
\begin{abstract}
Stablecoins support payments, trading, collateral, and cross-chain settlement across the digital-asset ecosystem. They also concentrate value behind issuer, custody, upgrade, oracle, and bridge keys while inheriting the quantum vulnerabilities of host-chain accounts, consensus, rollups, and privacy systems. This paper presents a Systematization of Knowledge (SoK) on post-quantum stablecoins. We develop a stablecoin-specific threat model, map cryptographic dependencies to monetary control surfaces, and classify migration choices along three dimensions: who can authorize a change, where the change must occur, and whether it is hybrid, post-quantum native, or an encapsulation of a classical component. We emphasize an authority--liability gap: the party able to migrate a component is often different from the holders, exchanges, protocols, or issuers that bear the loss if it fails. We review the relevant cryptographic primitives, but distinguish general blockchain failures from their stablecoin-specific effects. We also examine recent blockchain and issuer-controlled interoperability proposals, and relate migration choices to redemption, continuity, and intervention requirements under current stablecoin regulation. Our findings identify open problems in aggregate and threshold authorization, operation-specific security levels and costs, dormant and wrapped supply, private and compliance-enabled transfers, and measurement of quantum-vulnerable stablecoin exposure.
\end{abstract}

%% file: intro.tex
\section{Introduction}
\label{sec:introduction}

Stablecoins have become a critical settlement layer for the digital-asset economy. They are used for exchange liquidity, DeFi collateral, cross-border transfers, payment settlement, and increasingly as programmable cash-like instruments in tokenized finance. Recent SoK-style work emphasizes that their risk profile is not reducible to price stability: stablecoins combine payment-system design, legal obligations, governance, reserve management, smart-contract control, and cross-chain deployment choices~\cite{li2026stablecoinretail,ling2025stablecoinlego,zhang2025stablecoinsrwa}. This paper studies a dimension that remains under-systematized in that literature: the post-quantum security of stablecoin systems.

The problem is especially acute because stablecoin value is often concentrated behind a small number of accounts. A large fiat-backed issuer may rely on a relatively small set of treasury, minting, burning, pausing, blacklisting, upgrade, custody, oracle, and bridge keys. A crypto-collateralized stablecoin distributes some authority through governance, but still depends on oracle signatures, liquidation mechanisms, emergency controls, and the cryptography of the host chain. A bridged stablecoin adds lockboxes, relayers, committees, light clients, and cross-chain attestations. One successful key extraction can therefore affect supply, redemption, or liquidity far beyond the balance of an ordinary wallet.

Stablecoins also separate the actor able to correct a vulnerability from the parties that bear its cost. A host-chain community controls account and consensus changes; an issuer controls mint and freeze roles; a bridge operator controls a wrapped representation; holders, exchanges, lending protocols, and redeemers may absorb the resulting loss. We call this separation the \emph{authority--liability gap}. It is central to post-quantum stablecoin migration because a technically available upgrade may still be delayed or incomplete when authority, incentives, and legal obligations are distributed across different organizations.

Quantum computing changes the threat model for the public-key primitives used by blockchains. Shor's algorithm threatens RSA, ECDSA, Schnorr, EdDSA, BLS-style pairing-based signatures, and other constructions based on factoring or discrete logarithms, while Grover's algorithm reduces the security margin of symmetric search and hash-based primitives. Although the timeline for a cryptographically relevant quantum computer (CRQC) remains uncertain, recent resource estimates for secp256k1 distinguish at-rest and on-spend attacks and continue to refine the required logical-qubit and runtime estimates~\cite{babbush2026prx,luo2026pointadd}. This distinction matters for stablecoins because issuer, bridge, oracle, and custody keys are long-lived, repeatedly used, and easy to identify.

The operational urgency is not determined solely by the expected date of a CRQC. Migration is a multi-year process: standards must mature, libraries and HSMs must support new primitives, custodians and wallets must implement new formats, contracts may need precompiles or new verification paths, and governance processes must coordinate changes without disrupting redemption or settlement. General migration work therefore stresses cryptographic inventories, hybrid deployment, and crypto-agility~\cite{joseph2022pqmigrationIntro,nist2023nccoequantumreadiness,nistTransition2024,nistCryptoAgility2026}. For a stablecoin, however, a cryptographic inventory is also a financial-control inventory: every key or primitive maps to supply control, reserve movement, bridge liquidity, redemption, privacy, or governance.

Recent protocol work makes the coordination problem concrete. Zcash has proposed recoverability hooks for Orchard notes; Bitcoin proposals address new output types and the eventual treatment of legacy signatures; Ethereum proposals consider ML-DSA precompiles, native key delegation, and a broader execution/consensus/data-layer roadmap; issuer-controlled systems such as Circle's CCTP and xReserve expose cross-chain attestation and deprecation surfaces; and Circle's post-quantum roadmap sets out a phased issuer plan spanning contracts, custody, a new host chain, and account recovery~\cite{zcashzip2005,bip360,bip361,eip8051,eip8164,ethereumPQ2026,circleXReserve2025,circleCCTPDeprecation2025,circlePQRoadmap2026}. These proposals solve different parts of the problem. None by itself migrates the entire stablecoin stack.

\textbf{Problem statement and scope.}
This paper presents a Systematization of Knowledge (SoK) on \emph{post-quantum stablecoins}. We study how stablecoin systems can preserve minting authority, reserve management, transfers, bridge-backed supply, privacy, finality, and redemption against quantum-capable adversaries. Our focus is not a new cryptographic primitive or a prediction of the CRQC date. We instead systematize which stablecoin functions depend on which assumptions, who can migrate them, which losses propagate across organizational boundaries, and which protections can be deployed before host chains complete their own upgrades.

Several mechanisms studied here, including store-now-decrypt-later exposure, HSM migration, threshold authorization, and API security, also arise in other high-value financial systems. Stablecoins are nevertheless a useful exemplar because a single monetary claim composes issuer-controlled privileges with host-chain authentication and finality, bridges and oracles, custody and exchange infrastructure, and user-held keys. The actors able to migrate these components are often different from the parties that bear losses when they fail. We therefore treat stablecoins both as the primary object of study and as a concrete instance of multi-actor cryptographic migration in networked financial systems.

\textbf{Contributions.}
We make four contributions. First, we develop a stablecoin-centric threat model that maps quantum capabilities to issuer-key forgery, wrapped-supply inflation, oracle and governance manipulation, long-range history rewriting, proof failure, and retroactive disclosure. Second, we systematize cryptographic dependencies and control surfaces, including the authority--liability gap between migration authority and loss bearing. Third, we classify migration choices by control surface, deployment layer, and cryptographic strategy, and apply the taxonomy to current proposals and a staged migration plan. Fourth, we deepen the research agenda around missing stablecoin functionality, operation-specific security and cost, dormant and wrapped supply, private and compliance-enabled transfers, measurement, and regulatory continuity.

\textbf{Method.}
We reviewed peer-reviewed and archival research, standards and government guidance, blockchain improvement proposals and roadmaps, and issuer or open-source documentation. We use standards and peer-reviewed results for cryptographic claims; draft EIPs, BIPs, and ZIPs as evidence of active design directions; and issuer documentation to identify deployed control structures.

To make the coverage auditable at the category level, we organize the evidence around the stablecoin functions and control surfaces used throughout the paper: account authentication; mint/burn, treasury, custody, and upgrade authority; bridges and oracles; governance and consensus/finality; rollups and privacy systems; and encrypted operational channels. For each surface, we ask four questions: (i) which classical cryptographic assumption it depends on, (ii) which quantum capability breaks that assumption, (iii) which post-quantum replacement or migration mechanism is available, and (iv) which actor and deployment layer can actually perform the migration. Recent SoKs and surveys are used to identify adjacent categories and cross-check omissions, while primary sources are preferred for technical, standards, deployment, and regulatory claims that enter the taxonomy.

Our completeness claim is therefore category-level rather than bibliographic: we aim to cover the distinct stablecoin control surfaces, attack classes, and migration authorities, not every paper or implementation of each primitive. The paper is a systems SoK rather than a quantitative meta-analysis; newly proposed schemes can change the set of candidate mechanisms without necessarily changing the threat and migration categories.

\subsection{Related Work}
\label{sec:related-work}


\textbf{SoKs on blockchain systems.}
A large body of SoK-style blockchain work has developed taxonomies for adjacent parts of the ecosystem.
Layer-two and scaling SoKs classify payment channels, sidechains, rollups, sharding, and Bitcoin L2 designs~\cite{FC:GMRMG20,EPRINT:WSNH19b,albarat2024sokpublicsharding,qi2024bitcoinl2}.
Bridge-focused SoKs analyze cross-chain validation, lock-and-mint designs, light-client verification, bridge components, architectural flaws, and exploit classes~\cite{EPRINT:MBYS21,notland2024sokbridging}.
Security-focused SoKs on DeFi incidents, rug pulls, and decentralization provide incident taxonomies and measurement frameworks for smart-contract and governance risk~\cite{SP:ZXECWW23,zhang2022sokdecentralization,ACNS:OKMWK25}.
These works are essential for understanding the stablecoin environment, but they generally do not ask how quantum adversaries change the cryptographic assumptions of the control surfaces they classify.

\textbf{SoKs on stablecoins.}
Recent stablecoin SoKs study stablecoins as payment instruments, programmable settlement assets, and financial infrastructure.
The retail-payments SoK compares stablecoin payment arrangements with card networks and emphasizes differences in authorization, settlement, recourse, and user risk~\cite{li2026stablecoinretail}.
The Stablecoin LEGO SoK develops a design-risk framework based on stablecoin incidents and composability~\cite{ling2025stablecoinlego}.
Work on stablecoins and real-world asset tokenization develops taxonomies for design, metrics, and applications~\cite{zhang2025stablecoinsrwa}.
These studies provide the economic and architectural context for our work, but they do not systematize cryptographic migration, quantum attacks on issuer and bridge keys, or the post-quantum status of stablecoin-specific control paths.

\textbf{Post-quantum blockchains and primitives.}
Surveys review quantum-vulnerable blockchain primitives, post-quantum chain designs, migration strategies, and performance effects~\cite{fernandezcarames2020pqblockchain,yang2024pqquantumblockchains,ghosh2025quantumblockchain,baseri2025quantumrisk,mallick2025quantumdisruption}. Research on threshold and aggregate signatures, adaptor signatures, VRFs, commitments, and zero-knowledge proofs provides candidate components for custody, bridges, payment channels, consensus, and privacy~\cite{buser2023exoticsignatures,ESORICS:BDEKKL22,ICALP:AgrSteYad22,ProvSec:WanWu19,FC:TaiMorMaf21,ESORICS:EsgErsErk20}. Implementation work such as \emph{poqeth} shows that on-chain verifier design and opcode pricing can dominate native cryptographic performance~\cite{ASIACCS:KSKK25}. Our unit of analysis is different: we ask how these results specialize to stablecoin control, claims, and loss propagation.

\textbf{Migration proposals and chain-specific readiness.}
Protocol proposals cover complementary pieces of migration. ZIP-2005 adds pre-crisis recoverability for Zcash Orchard notes~\cite{zcashzip2005}; EdDSA-chain work preserves deployed address formats while registering post-quantum credentials~\cite{EPRINT:BalChaRoy25}; Bitcoin BIP-360 and BIP-361 address new output semantics and legacy-signature sunset policy~\cite{bip360,bip361}; and Ethereum EIP-8051 and EIP-8164 propose specialized ML-DSA verification and native EOA key delegation~\cite{eip8051,eip8164}. Other ecosystems provide useful migration case studies. 
Sui inventories exposure in account signatures, BLS-based randomness,
TLS, bridges, and zkLogin~\cite{suiQuantum2025}. The existing zkLogin
construction uses a Groth16 proof over an OIDC credential~\cite{CCS:BCJLMR24},
while Sui's PQ migration analysis proposes replacing both the credential
signature and proof system while retaining the account
address~\cite{suiQuantum2025}.
Solana reports initial Falcon implementations by Anza and Firedancer and a staged path from new PQ wallets to legacy-wallet migration, while an application-level Winternitz vault demonstrates an overlay approach~\cite{solanaQuantum2026}. Coinbase's 2026 Quantum Advisory Council reports separately emphasize cross-chain migration and the unresolved governance of assets that never migrate~\cite{coinbaseQuantumCouncil2026,coinbaseAbandonedCoins2026}. Circle's post-quantum roadmap phases readiness, transition, and switch, and states which components the issuer controls and which it must leave to host chains, vendors, and contract owners~\cite{circlePQRoadmap2026}. We compare the principal protocol and chain proposals in Section~\ref{sec:proposals}, emphasizing which stablecoin components remain outside each proposal's scope.

%% file: background.tex
\section{Background}
\label{sec:background}

This section introduces the stablecoin designs, control relationships, regulatory constraints, and cryptographic building blocks used throughout the paper. We provide generic formal definitions in Appendix~\ref{app:definitions}.

\subsection{Stablecoin Designs and Control Surfaces}

Stablecoins are crypto-assets designed to maintain a relatively stable value, typically by referencing an external asset, a basket of assets, or an endogenous stabilization mechanism. For post-quantum analysis, the important distinction is not only how the peg is maintained, but which entities control issuance, redemption, collateral, and administrative functions, and which cryptographic keys authorize those actions.

\pparagraph{Fiat-backed stablecoins}
Fiat-backed stablecoins, such as USDT, USDC, PYUSD, and EURC, are issued by centralized entities that hold off-chain reserves and promise redemption at a fixed reference value. On-chain, they are commonly implemented as upgradeable contracts with privileged roles for minting, burning, pausing, blacklisting, contract upgrades, or treasury movements. Circle's public EVM implementation, for example, separates owner, proxy-owner, master-minter, pauser, and blacklister roles~\cite{circleStablecoinEVM}. These roles are controlled by issuer keys, multisignature wallets, custody providers, MPC policies, or HSMs. Quantum risk is therefore concentrated in a small number of high-value control accounts.

\pparagraph{Crypto-collateralized stablecoins}
Crypto-collateralized stablecoins, such as DAI and LUSD, are backed by over-collateralized positions in volatile crypto-assets locked in smart contracts. Their security depends on contract correctness, liquidation mechanisms, oracle integrity, governance, keeper infrastructure, and the cryptography of the host chain. They inherit user- and validator-key risk from the underlying chain while adding protocol-specific exposure through oracle signers, governance keys, emergency shutdown mechanisms, and collateral contracts.

\pparagraph{Algorithmic and hybrid designs}
Algorithmic stablecoins attempt to maintain a peg through supply expansion and contraction, arbitrage incentives, seigniorage-style mechanisms, or multi-token structures. Historical examples include Basis Cash and TerraUSD (UST), while hybrid systems such as FRAX combined collateral with algorithmic mechanisms. These systems often rely heavily on governance keys, oracle inputs, liquidity-management contracts, and upgrade paths. Quantum compromise of those controls can destabilize the peg without directly stealing reserve assets.

\pparagraph{Bridged and wrapped stablecoins}
Stablecoins are deployed across chains through native issuance, canonical bridges, third-party bridges, and wrapped representations. In a lock-and-mint design, tokens are escrowed on a source chain and a corresponding representation is minted on a destination chain after a committee, validator set, oracle, or light client authenticates the source event. Burn-and-mint systems authorize issuance or release after a source-chain burn. The stablecoin therefore depends on bridge signers, message formats, source-chain finality, relayers, lockboxes, and destination contracts. A forged attestation can create unbacked wrapped supply or release locked value.

\subsection{Control, Settlement, and Claims}
\label{sec:control-model}

A stablecoin is not a single contract. We model it as three coupled layers. The \emph{control plane} changes supply or the rules governing it: mint and burn keys, treasury operations, pause and blacklist roles, proxy upgrades, governance, oracle administration, and bridge attestation. The \emph{settlement substrate} authenticates users and orders transactions: host-chain accounts, validator or finality signatures, data-availability commitments, rollup proof systems, sequencers, light clients, and cross-chain verification. The \emph{claim and distribution layer} contains holders, custodians, exchanges, DeFi protocols, payment applications, redemption channels, and wrapped representations. Parties in the last layer often absorb losses even when they do not control the failed component.

Table~\ref{tab:authority-liability} makes the resulting authority--liability gap explicit. A technically available migration may still fail because the actor able to deploy it lacks the incentive, legal authority, or coordination channel to protect the principal loss bearer.

\begin{table*}[t]
\centering
\scriptsize
\setlength{\tabcolsep}{3.5pt}
\begin{tabularx}{\textwidth}{@{}p{2.4cm}p{3.0cm}p{2.55cm}p{2.9cm}Y@{}}
\toprule
\textbf{Control surface} & \textbf{Stablecoin consequence} & \textbf{Actor able to migrate} & \textbf{Principal loss bearer} & \textbf{Dependency that may block migration} \\
\midrule
Mint, burn, treasury, upgrade & Unauthorized supply, reserve movement, freeze, or contract replacement & Issuer, DAO, custodian & Holders, redeemers, issuer & Host-chain verifier, HSM/MPC support, governance delay \\
Immutable contracts verifying \texttt{ecrecover} signatures & Forged permits and gasless transfers into unupgradeable DeFi, custody, and payment contracts & Host chain (hard fork replacing \texttt{ecrecover}) & Contract owners and their users & No upgrade path; 65-byte signature ABI; PQ keys require a registry~\cite{circlePQRoadmap2026} \\
Oracle and liquidation & False collateral state, liquidations, or peg failure & Oracle network, protocol governance & Vault users, liquidity providers, holders & Aggregation, update frequency, contract gas \\
Bridge lockbox and attestation & Unbacked wrapped supply or release of locked assets & Bridge operator, issuer, source/destination chains & Wrapped-token holders, exchanges, DeFi protocols & Source finality, committee signatures, redemption coordination \\
Host-chain accounts and consensus & User theft, validator forgery, history rewriting & Chain governance and client implementers & Users and all applications on the chain & Hard fork, wallet support, light clients, dormant accounts \\
Rollup and privacy systems & Invalid state transitions, false compliance proofs, retroactive disclosure & Rollup or privacy-system operator & Users, issuers, regulated intermediaries & Proof replacement, recursion, wallet proving cost \\
Dormant and inaccessible balances & Persistent exposed supply or coercive freeze/recovery & User, issuer, or chain depending on design & Dormant holder; active holders through confidence loss & Lost keys, legal ownership, recovery proof, deadline policy \\
\bottomrule
\end{tabularx}
\caption{Stablecoin control surfaces separate migration authority from financial liability.}
\label{tab:authority-liability}
\end{table*}

\begin{figure*}[t]
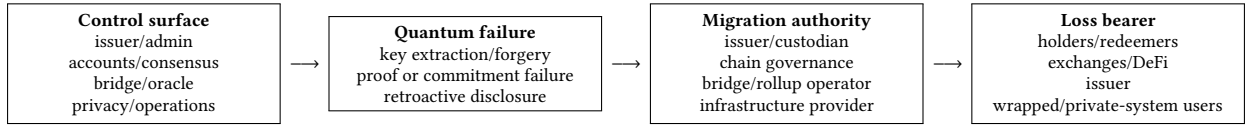

\centering
\footnotesize
\begin{tabular}{@{}c@{\hspace{0.45em}$\longrightarrow$\hspace{0.45em}}c@{\hspace{0.45em}$\longrightarrow$\hspace{0.45em}}c@{\hspace{0.45em}$\longrightarrow$\hspace{0.45em}}c@{}}
\fbox{\parbox{0.19\textwidth}{\centering\textbf{Control surface}\\issuer/admin\\accounts/consensus\\bridge/oracle\\privacy/operations}} &
\fbox{\parbox{0.19\textwidth}{\centering\textbf{Quantum failure}\\key extraction/forgery\\proof or commitment failure\\retroactive disclosure}} &
\fbox{\parbox{0.19\textwidth}{\centering\textbf{Migration authority}\\issuer/custodian\\chain governance\\bridge/rollup operator\\infrastructure provider}} &
\fbox{\parbox{0.19\textwidth}{\centering\textbf{Loss bearer}\\holders/redeemers\\exchanges/DeFi\\issuer\\wrapped/private-system users}}
\end{tabular}
\caption{Stablecoin quantum risk is compositional: a cryptographic failure propagates through a control surface, but the actor able to migrate that surface and the party bearing the resulting loss may differ.}
\label{fig:authority-liability-flow}
\end{figure*}

\subsection{Regulatory Constraints on Migration}
\label{sec:regulation}

Regulation does not select a post-quantum algorithm, but it constrains acceptable migration behavior. MiCA requires governance and ICT controls, business continuity, recovery and orderly-redemption plans, and par redemption for e-money tokens; relevant obligations also apply to significant e-money tokens through Article~55~\cite{mica2023}. In the United States, the GENIUS Act requires one-to-one reserves and disclosed redemption procedures, directs regulators to establish operational and information-technology risk standards, and requires issuers to retain the technical capability to comply with lawful orders that may seize, freeze, burn, or prevent transfers~\cite{geniusActText2025}. The OCC's February 2026 proposal addresses redemption, risk management, custody, audits, and operational backstops~\cite{occGenius2026}; a June 2026 proposal separately addresses BSA/AML and sanctions programs~\cite{occGeniusAml2026}.

\begin{table}[t]
\centering
\scriptsize
\setlength{\tabcolsep}{3pt}
\begin{tabularx}{\columnwidth}{@{}p{1.55cm}p{2.05cm}Y@{}}
\toprule
\textbf{Source} & \textbf{Requirement} & \textbf{Migration consequence} \\
\midrule
MiCA Art. 34 & Governance, ICT controls, continuity & Key and contract migration must be inventoried, tested, and auditable. \\
MiCA Arts. 46--47, 49, 55 & Recovery, orderly redemption, par redeemability & Freezes, reissuance, and claim contracts must preserve timely and equitable redemption. \\
GENIUS Act \S4 & Reserves, disclosed redemption, operational/IT risk & A PQ incident is an operational-risk and disclosure event, not only a protocol bug. \\
GENIUS Act \S4(a)(6) and OCC proposals & Lawful orders, custody, audit, AML/sanctions & Migration must preserve required freeze, burn, screening, blocking, and reporting controls. \\
\bottomrule
\end{tabularx}
\caption{Examples of regulatory duties that constrain stablecoin migration.}
\label{tab:regulatory}
\end{table}

These provisions sharpen the stablecoin-specific problem. A chain-level proposal that abandons old accounts may conflict with an issuer's redemption duties. Conversely, a migration that preserves user signatures but removes pause, blacklist, recovery, or lawful-order capability may be unacceptable for a regulated issuer. Regulation therefore constrains recovery semantics, authority, audit evidence, and continuity rather than merely providing external motivation.

\subsection{Cryptographic Primitives, Quantum Vulnerabilities, and Post-Quantum Candidates}
\label{sec:crypto-basic}

\pparagraph{Hash functions}
Hash functions support transaction identifiers, Merkle trees, state commitments, nullifiers, proof systems, and address derivation. Grover's algorithm gives a quadratic preimage speedup, while BHT gives a stronger asymptotic collision attack under quantum-memory assumptions~\cite{STOC:Grover96,brassard1997quantum}. Grover parallelizes poorly, so 256-bit hashes retain roughly a 128-bit generic post-quantum margin. Hash-based signatures such as SLH-DSA, XMSS, and LMS, and transparent STARK systems, inherit security primarily from these properties~\cite{NISTPQC:SPHINCS+22,PQCRYPTO:BucDahHul11,EPRINT:KamFlu17,EPRINT:BBHR18}. Stablecoin systems should therefore prioritize public-key migration rather than replace well-sized hash functions indiscriminately.

\pparagraph{Digital signatures}
ECDSA, Schnorr, EdDSA, RSA-PSS, and BLS signatures rely on factoring, discrete-logarithm, or pairing assumptions and fall to Shor's algorithm~\cite{JC:Schnorr91,JCEng:BDLSY12,AC:BonLynSha01,FOCS:Shor94}. Once a public key is exposed, a CRQC can recover the secret and create signatures indistinguishable from honest ones. This threatens not only future transactions but also the evidentiary value of historical signatures, reserve attestations, and audit logs. Stablecoin mint, treasury, bridge, oracle, and upgrade keys are typically in the at-rest regime because their public keys are long-lived and visible~\cite{babbush2026prx}.

NIST has finalized ML-DSA and SLH-DSA and is continuing standardization of FN-DSA, based on Falcon~\cite{nistFips204,nistFips205,nistPQCProject2026}. ML-DSA has kilobyte-scale public keys and signatures and is a plausible default for low-frequency issuer administration. SLH-DSA offers conservative hash-based security at much larger signature sizes. Falcon/FN-DSA is more compact but has demanding implementation and side-channel requirements. For tightly controlled low-frequency keys, NIST SP~800-208 also approves the stateful hash-based LMS and XMSS families and stresses careful state management and hardware-backed key handling; this profile is better matched to treasury or recovery controls than to retail transfers~\cite{nistSP800208}. In May 2026, NIST advanced nine additional signature candidates, FAEST, HAWK, MAYO, MQOM, QR-UOV, SDitH, SNOVA, SQIsign, and UOV, to a third evaluation round~\cite{nistAdditionalSigs2026}. None of these developments provides a drop-in replacement for BLS-style non-interactive aggregation.

\pparagraph{Multisignatures and threshold signatures}
Stablecoin custody, issuer administration, exchange wallets, bridge committees, and validator sets rely on distributed authorization. Threshold ECDSA, FROST-style Schnorr, and threshold BLS remain vulnerable because Shor recovers the aggregate secret from the aggregate public key; the attacker need not compromise each share~\cite{CCS:GenGol18,SAC:KomGol20,CCS:BacLos22}. Lattice multisignatures and threshold signatures exist~\cite{CANS:ElBStu16,ICALP:AgrSteYad22,AC:ChaTesZhu24}, and recent work specifically targets threshold ML-DSA through MPC and identifiable-abort constructions~\cite{EPRINT:BCEPT25,EPRINT:BCdENP25}. These results narrow the primitive gap, but stablecoin deployment still requires dynamic committees, signer accountability, HSM/MPC support, key rotation and recovery, compact verification, and audit evidence.

\pparagraph{Key exchange and authenticated channels}
Issuer--custodian APIs, reserve operations, bank and exchange connections, oracle feeds, node links, and incident-response channels depend on encrypted communication. DH/ECDH, ECIES, and RSA-OAEP lose their security under Shor~\cite{IMA:Smart01}. This creates store-now-decrypt-later exposure for archived operational and customer data. ML-KEM is standardized, and HQC has been selected for ongoing standardization as a backup KEM~\cite{nistFips203,nistHQC2025}. Hybrid KEMs can therefore be deployed in issuer infrastructure before host-chain account formats change. An active IETF standards-track Internet-Draft defines X25519MLKEM768, SecP256r1MLKEM768, and SecP384r1MLKEM1024 for TLS~1.3, providing a concrete hybrid migration path for issuer--custodian and issuer--bank channels while the document remains work in progress~\cite{ietfTLSHybridMLKEM2026}.

\pparagraph{Randomness, VRFs, and leader election}
VRFs and randomness beacons support validator or committee selection, auctions, and some governance and liquidation mechanisms. EC-VRFs and threshold-BLS beacons inherit discrete-log or pairing risk; a recovered key makes future outputs predictable and can permit bias or control~\cite{FOCS:MicRabVad99,CCS:NTSL25}. Symmetric-primitive and lattice-based post-quantum VRFs have been proposed~\cite{ESORICS:BDEKKL22,EPRINT:Abraham18}, but proofs are larger and integration with high-throughput consensus and light-client sampling remains immature.

\subsubsection{Privacy and Advanced Functionality}

\pparagraph{Zero-knowledge proofs}
Groth16, PLONK with KZG, Bulletproofs, and Halo-style systems derive soundness from pairing or discrete-log assumptions; Shor can therefore enable accepting proofs for false statements, including invalid rollup transitions or creation of unbacked shielded balances~\cite{EC:Groth16,EPRINT:GabWilCio19,SP:BBBPWM18,C:BDFG21,EPRINT:SHJKGA23}. STARKs rely primarily on hash commitments and FRI-style testing and are treated as post-quantum under appropriate hash parameters~\cite{EPRINT:BBHR18}. Lattice-based SNARKs and zero-knowledge arguments, and MPC-in-the-head systems such as Ligero, Aurora, and Picnic, provide alternatives with different proof-size, recursion, and verifier-cost trade-offs~\cite{C:ACLMT22,C:LyuNguPla22,C:NguSei22,C:ESLL19,PKC:GZGPX25,CCS:AHIV17,EC:BCRSVW19,CCS:CDGORR17}. These costs matter directly for private stablecoin transfers, reserve proofs, compliance proofs, and rollup settlement.

\pparagraph{Commitments}
Pedersen commitments are statistically hiding but computationally binding under the discrete-log assumption; Shor breaks binding and permits inconsistent openings~\cite{C:Pedersen91}. This can invalidate confidential balances or reserve proofs without compromising an issuer signing key. KZG commitments similarly inherit pairing risk, while hash-based commitments retain a post-quantum margin at 256-bit parameters. Lattice commitments integrate with lattice proof systems but have different proof and verifier costs~\cite{C:AttLyuSei20}. The relevant migration question is therefore the entire proof-and-commitment stack, not the commitment in isolation.

\paragraph*{Encryption and AEAD}
Symmetric encryption and AEAD face generic Grover-style key search; 256-bit keys provide a practical post-quantum baseline. Under superposition access to a keyed primitive, Simon-style attacks can break some MAC and mode constructions, although such access is unrealistic for ordinary stablecoin attackers~\cite{C:KLLN16}. Public-key encryption based on ECIES or RSA falls under Shor. A post-quantum design should therefore use ML-KEM or a hybrid KEM with 256-bit AEAD and should consider forward secrecy to reduce the value of archived traffic~\cite{C:BelNgTac19}.

\pparagraph{Exotic signatures (ring, blind, adaptor, and aggregate)}
\label{sec:exotic}
These appear in stablecoin privacy layers, bridge attestations, payment
channels, and atomic swaps.
 
\noindent \emph{Ring signatures}~\cite{AC:RivShaTau01} let a signer sign on behalf of
a ring $R = \{\pk_1, \ldots, \pk_n\}$ with \textbf{unforgeability} and
\textbf{anonymity} (the signer's identity is hidden within $R$). DL-based
constructions (LSAG, MLSAG in Monero) lose both under Shor: once any ring
member's $\sk$ is recoverable, the adversary can both forge and deanonymize,
implying \emph{retroactive deanonymization} of historical transactions and
any ring-based stablecoin pool. PQ candidates: lattice ring signatures
(Raptor~\cite{ACNS:LuAuZha19}, lattice RingCT~\cite{ACISP:TSSLKB18,ACISP:TKSSLC19}) and
code-based variants.
 
\noindent \emph{Blind signatures} (e-cash, confidential withdrawals) require
\textbf{blindness} and \textbf{one-more unforgeability}. For DL-based
schemes, Shor enables recovery of the signing key and therefore destroys
unforgeability. The effect on blindness is construction-dependent: schemes
with statistical blindness can retain that property even after the signing
key becomes recoverable. Lattice blind signatures~\cite{CCS:AKSY22,CCS:BLNS23}
are available but with sizes far above ECDSA-based variants.
 
\noindent \emph{Adaptor signatures}~\cite{AC:AEEFHM21} tie signature publication to
revealing a witness $w$ for $Y = f(w)$, requiring \textbf{pre-signature
adaptability}, \textbf{witness extractability}, and \textbf{aEUF-CMA}. In
EC-based constructions $w$ is a discrete log, so Shor extracts $w$ directly
from $Y = g^w$ without ever observing a signature; breaking the entire
construction. PQ candidates target payment-channel
networks~\cite{ESORICS:EsgErsErk20,FC:TaiMorMaf21}.
 
\noindent \emph{Aggregate signatures.} BLS~\cite{EC:BGLS03,AC:BonLynSha01} compresses
$n$ signatures over distinct messages and keys into a single 48-byte
element; all variants are broken by Shor. Lattice sequential aggregate
signatures~\cite{ProvSec:WanWu19} exist but are limited in non-interactive
aggregation and rogue-key resistance. This is among the most pressing open
problems for post-quantum bridge attestations.

%% file: threats.tex
\section{Threat Model and Quantum Attack Surface}
\label{sec:threats}

This section identifies quantum-relevant threats to stablecoin systems.
We distinguish \emph{capabilities} (what the adversary can compute), \emph{attack patterns} (how those capabilities are used), and \emph{stablecoin control surfaces} (which assets or operations are affected).

\subsection{Adversary Model and Capabilities}
\label{sec:adversary}

We formalize the adversary in three layers: classical capabilities, quantum capabilities, and the time horizon over which capabilities accumulate.

\pparagraph{Classical adversary}
Let $\adv_{\mathrm{cl}}$ denote a $\PPT$ adversary. We assume $\adv_{\mathrm{cl}}$ has the following abilities, all of which are realistic for present-day attackers on public blockchains:
\begin{itemize}[leftmargin=2em]
    \item \textbf{Public-ledger observation}: read all on-chain state, including transactions, contract code, public keys, signatures, Merkle proofs, rollup state commitments, bridge attestations, and oracle messages.
    \item \textbf{Transaction submission}: post arbitrary transactions; perform front-running, sandwiching, and MEV; deploy contracts.
    \item \textbf{Node and P2P operations}: run validators, RPC providers, relayers; observe and archive mempool traffic and gossip metadata.
    \item \textbf{Network archival}: passively record encrypted off-chain traffic (TLS sessions, custodian APIs, oracle feeds, bridge relayer traffic). This is the substrate for store-now-decrypt-later attacks. For off-chain SNDL, the adversary needs a ciphertext-recording vantage point, such as an ISP or transit path, enterprise or cloud-network infrastructure, a compromised network appliance, or service-side packet logging; endpoint compromise is a stronger capability and is not assumed. Public on-chain ciphertexts require no privileged vantage point because any observer can archive them.
    \item \textbf{Selective corruption}: compromise or bribe a bounded subset of off-chain participants (bridge committee members, oracle signers, custodians, validator operators). We do not assume any single party is fully corrupted.
\end{itemize}

\pparagraph{Quantum adversary}
The quantum adversary $\adv_{q}$ is a $\QPT$ machine that additionally holds (or has time-shared access to) a CRQC sufficient for the workloads below.
\begin{itemize}[leftmargin=2em]
    \item \textbf{Shor's algorithm}~\cite{FOCS:Shor94}: given a target public key $\pk \in \mathbb{G}$ for any blockchain-relevant group $\mathbb{G}$ (secp256k1, secp256r1, Curve25519, BLS12-381), recover the corresponding secret key $\sk$ in time $\mathrm{poly}(\log |\mathbb{G}|)$. Recent circuit estimates put the logical-qubit requirement for 256-bit prime-field ECDLP at roughly $10^3$. For secp256k1, Babbush et al.~\cite{babbush2026prx} report time--space trade-offs using at most about 1,200 logical qubits and 90 million Toffoli gates, or about 1,450 logical qubits and 70 million Toffoli gates. More recent work by Luo et al.~\cite{luo2026pointadd} gives an 835-logical-qubit construction for a 256-bit prime-field curve. These are \emph{logical} resource counts; the number of physical qubits required for a fault-tolerant implementation is substantially larger and depends on the error-correction architecture, physical error rate, and target runtime.
    \item \textbf{Grover's algorithm}~\cite{STOC:Grover96}: $\Theta(2^{n/2})$ queries to invert an arbitrary $n$-bit function. 
    Parallel speedup is only $\sqrt{k}$ in the number of parallel oracles, which substantially limits Grover's threat at 256-bit hash output sizes.
    \item \textbf{Simon's algorithm / quantum chosen-input attacks}~\cite{simon1997power}: relevant against some MACs and stream ciphers under \emph{superposition} queries to the keyed primitive. In practice, blockchain attackers cannot obtain such queries against issuer infrastructure, so the threat is mostly theoretical for stablecoins.
\end{itemize}

\begin{remark}[Granularity of quantum compromise]
\label{rem:granularity}
Quantum capability is not all-or-nothing. We distinguish three regimes:
\begin{enumerate}[leftmargin=2em]
    \item \textbf{Targeted offline key extraction}: the adversary can afford one Shor run per high-value target. This already destroys treasury and bridge keys whose public keys are static and known.
    \item \textbf{Batched extraction}: the adversary can run Shor against many keys per day. All exposed EOAs become extractable; the bottleneck shifts to ranking targets by value.
    \item \textbf{Online race-the-mempool}: the adversary intercepts a transaction at broadcast time, extracts $\sk$ from the revealed $\pk$, and submits a competing transaction before the original confirms. Even short windows (12--15s on Ethereum) become feasible once Shor runtimes drop into the seconds range.
\end{enumerate}
For stablecoins, the at-rest regime is already devastating because issuer, bridge, and oracle keys live in static, on-chain addresses; the on-spend regime is the regime ordinary user wallets need to worry about.
\end{remark}

\pparagraph{Time horizon}
A useful planning model is \emph{Mosca's inequality}~\cite{mosca2018cybersecurity}: a cryptosystem must be considered insecure if $X + Y > Z$, where $X$ is the shelf-life of the protected data, $Y$ is the migration time, and $Z$ is the time until a CRQC is operational. For stablecoins:
\begin{itemize}[leftmargin=2em]
    \item $X$ is years to decades: reserve attestations, compliance records, and shielded transaction histories may need to remain unforgeable or private for very long horizons.
    \item $Y$ is large because stablecoins span issuer infrastructure, smart contracts, bridges, rollups, custodians, exchanges, wallets, and host-chain consensus, and migration of each layer requires its own governance process.
    \item $Z$ is uncertain but trending downward: recent resource estimates~\cite{babbush2026prx,luo2026pointadd} continue to lower the qubit and time cost of such attacks.
\end{itemize}

\subsection{Stablecoin Attack Patterns}
\label{sec:attack-patterns}

We now systematize the attack patterns that follow from the capabilities of Section~\ref{sec:adversary}. Each pattern is stated in terms of the target property, attacker capability, goal, and  impact.

\begin{attack}[At-rest key extraction and forgery]
\label{atk:at-rest}
\textbf{Target property}: EUF-CMA security of the deployed signature scheme. \textbf{Capability}: Shor on already-revealed $\pk$. \textbf{Goal}: produce signatures $\sigma$ on attacker-chosen messages $m$ authenticated by $\sk$, without ever observing legitimate signatures.
\end{attack}

\noindent The set of \emph{at-rest exposed} stablecoin keys typically includes treasury and mint multi-sig signer keys; proxy-admin / upgrade keys; pause and blacklist authorities; bridge committee keys; oracle signer keys; custodian co-signing keys; and host-chain validator keys. Each is a \emph{concentrated control surface}: one forged signature may mint supply, freeze balances, replace implementation logic, push a malicious oracle update, or finalize a malicious checkpoint.

\begin{attack}[Key extraction race (on-spend)]
\label{atk:on-spend}
\textbf{Target}: account-level signatures whose $\pk$ is revealed at spending time. \textbf{Capability}: fast Shor extraction and mempool monitoring. \textbf{Goal}: extract $\sk$ from the broadcast $\pk$ and submit a higher-fee replacement transaction before the original confirms.
\end{attack}

\noindent For stablecoins, this primarily threatens user holdings. The system-level impact is bounded by chain throughput, but the individual-user impact is total loss.

\begin{attack}[Retroactive disclosure - Store-now decrypt-later (SNDL) and privacy loss]
\label{atk:retro}
\textbf{Target property}: confidentiality of \emph{archived ciphertexts}
(IND-CCA / IND-CPA of past KEM/PKE traffic) \emph{and} anonymity / hiding of
\emph{on-chain privacy mechanisms} (ring signatures, shielded notes, viewing
keys, ZK witness-hiding).
\textbf{Capability}: archive ciphertexts and on-chain artifacts now; later
apply Shor to the underlying DLP / pairing / factoring assumption.
\textbf{Goal}: recover plaintexts of past communications and reconstruct the
historical transaction graph of past private transfers.
\end{attack}

For the off-chain branch, passive recording therefore requires network visibility but not plaintext or endpoint access; for the on-chain branch, ordinary public-ledger access is sufficient to archive the relevant artifacts.
 
\noindent This pattern unifies two related threats that share an ``archive now, break
later'' structure but affect different surfaces. \emph{Off-chain SNDL} on
encrypted issuer--custodian APIs, redemption queues, and compliance traffic
exposes signing-ceremony schedules, reserve-rebalancing instructions and bank
references, KYC envelopes, and bridge relayer topology, enabling timed
treasury attacks, phishing of banking partners, regulatory harm, and targeted
compromise of bridge committees, all without any on-chain action.
\emph{On-chain retroactive privacy loss} affects stablecoins routed through
ring-based pools (Monero-style), shielded protocols (Zcash, Aztec,
Railgun), or privacy-preserving rollups: once the underlying assumption
falls, payer--payee links, amounts, timing, and DeFi positions become
reconstructible years after the transactions appeared private. The harm in
both sub-cases is informational and \emph{cannot be repaired by re-issuance
or freezing}; the data is already out, and the cryptographic break is
permanent. Privacy mechanisms also differ in their assumptions: the Circle Research--Inco confidential-token design uses fully homomorphic encryption over balances and amounts, whereas note-based systems combine commitments, zero-knowledge proofs, and encrypted note data~\cite{circleInco2024}. The assessment must therefore inventory the actual privacy construction rather than infer security from the token interface.

\begin{attack}[Bridge and wrapped-asset forgery]
\label{atk:bridge}
\textbf{Target property}: EUF-CMA or threshold-unforgeability of the bridge attestation scheme. \textbf{Capability}: Shor against committee public keys or aggregate BLS keys. \textbf{Goal}: produce attestations for lock/burn events never occurred.
\end{attack}

\noindent A successful forgery on a lock-and-mint bridge mints unbacked wrapped stablecoin on the destination chain; on a burn-and-mint bridge it releases native value without a corresponding burn. Because wrapped assets are often traded as fungible claims, the attack can propagate through exchanges, AMMs and lending platforms before reconciliation detects the mismatch. Circle xReserve illustrates the dependency structure of a modern issuer-backed interoperability system: a Circle-deployed reserve contract, an API-issued attestation, and partner-chain mint/burn logic jointly determine issuance and redemption~\cite{circleXReserve2025}. A post-quantum migration must cover all three rather than only the destination token.

\begin{attack}[Oracle and governance manipulation]
\label{atk:oracle}
\textbf{Target property}: signature unforgeability of oracle reports or governance votes; ZK soundness for off-chain attested data; VRF uniqueness/pseudorandomness for auctions or committee sampling. \textbf{Capability}: Shor on signer/voter keys or on pairing assumptions. \textbf{Goal}: induce false price updates, unjustified liquidations, or replacement of oracle/governance sets.
\end{attack}

\noindent For crypto-collateralized and hybrid stablecoins, oracle and governance forgery can trigger mass liquidations, conceal undercollateralization, or replace peg-maintenance modules.

\begin{attack}[Commitment equivocation in confidential balances]
\label{atk:equiv}
\textbf{Target property}: computational binding of a Pedersen, KZG, or related commitment used for confidential balances or proofs of solvency. \textbf{Capability}: Shor on the underlying DLP or pairing assumption. \textbf{Goal}: open the same commitment to inconsistent values.
\end{attack}

This pattern does not require a signing key. A DLOG break can let proofs of solvency overstate assets, or let confidential-balance systems equivocate about balances, whenever binding relies on discrete-log assumptions.

\begin{attack}[Soundness break in validity proofs]
\label{atk:zk-snd}
\textbf{Target property}: computational soundness of pairing-based or DLP-based SNARKs. \textbf{Capability}: Shor against the soundness assumption. \textbf{Goal}: produce accepting proofs $\pi$ for false statements $x \notin \mathcal{L}$.
\end{attack}

\noindent Direct impacts include invalid rollup state transitions, invalid shielded-pool balance updates, and forgeable compliance or proof-of-reserves proofs.

\pparagraph{Summary of attack patterns}
Table~\ref{tab:patterns}
summarizes
the patterns, the security property they target, and the dominant stablecoin
control surface they affect. 


\begin{table*}[!t]
\centering
\small
\setlength{\tabcolsep}{4pt}
\begin{tabularx}{\textwidth}{@{}p{2.6cm}p{4.2cm}Y@{}}
\toprule
\textbf{Pattern} & \textbf{Property broken} & \textbf{Stablecoin surface} \\
\midrule
A. At-rest extraction      & EUF-CMA of signature            & Mint / treasury / admin / oracle / bridge keys \\
B. On-spend race           & EUF-CMA at key disclosure               & User wallets and dormant balances \\
C. Retroactive disclosure      & IND-CCA of past KEM/PKE; anonymity / hiding of privacy schemes (offline) & Custody / compliance / API traffic; shielded stablecoin history \\
D. Bridge forgery          & Threshold / aggregate unforgeability   & Wrapped supply and lockboxes \\
E. Oracle / governance     & Report/vote unforgeability           & Peg logic, liquidation, parameter updates \\
F. Commitment equivocation & Commitment binding                    & Confidential balances, proof-of-reserves \\
G. ZK soundness            & Proof-system soundness    & Rollups, shielded pools, compliance proofs \\
\bottomrule
\end{tabularx}
\caption{Quantum-enabled attack patterns for stablecoin systems.}
\label{tab:patterns}
\end{table*}

\subsection{Layered Composition of Threats}
\label{sec:layered-threats}

The attacks above do not occur in isolation: the same cryptographic break can
appear at different layers of the stablecoin stack. We therefore summarize how
the labeled attacks compose across the stablecoin layer stack.

\pparagraph{Application layer}
At the stablecoin-application layer, the dominant risks are concentrated control
surfaces: mint, burn, pause, blacklist, treasury, upgrade, oracle, governance,
custody, and bridge keys. The primary attack is key extraction and signature
forgery against already exposed high-value public keys
(Attack~\ref{atk:at-rest}).
Bridge-controlled stablecoin supply is exposed through forged bridge
attestations and wrapped-asset inflation
(Attack~\ref{atk:bridge}), while
oracle and governance manipulation affects collateral ratios, liquidations, peg
modules, and parameter updates
(Attack~\ref{atk:oracle}). If
the stablecoin or its surrounding infrastructure uses confidential balances,
proof-of-reserves commitments, shielded transfers, or compliance proofs, then
commitment equivocation and proof-system soundness attacks also become
application-layer concerns (Attacks~\ref{atk:equiv}
and~\ref{atk:zk-snd}).
A further application-layer surface is the \texttt{ecrecover}-based
verification of off-chain authorizations such as EIP-2612 permits and EIP-3009
transfers. The stablecoin contract itself may be upgradeable, but most contracts
that consume these signatures are not, so Circle's post-quantum roadmap
identifies a host-chain replacement of \texttt{ecrecover} as the only remedy and
places remediation responsibility on contract
owners~\cite{circlePQRoadmap2026}.

\pparagraph{Blockchain stack}
At the blockchain stack layer, stablecoins inherit account authentication, validator signatures, finality, light-client verification, bridge verification, and rollup settlement from their host chains. User balances are exposed through on-spend key extraction when a public key is revealed only at transaction time (Attack~\ref{atk:on-spend}), while validators, sequencers, light-client committees, and exposed account or contract-admin keys fall under at-rest key extraction (Attack~\ref{atk:at-rest}). Long-range proof-of-stake rewriting is not a separate primitive attack; it is a composition of at-rest key extraction against historical validator keys with the chain's finality and weak-subjectivity assumptions. Similarly, rollup or bridge failure may arise from forged attestations (Attack~\ref{atk:bridge}) or from a soundness break in pairing- or DLP-based validity proofs (Attack~\ref{atk:zk-snd}). Thus, a stablecoin contract can use PQ-safe issuer controls while still depending on quantum-vulnerable host-chain accounts, validator signatures, light-client proofs, or rollup verifiers.

\pparagraph{Network, metadata, and privacy layer}
Stablecoin operators also depend on off-chain encrypted channels, APIs, custody procedures, banking partners, customer-support systems, compliance databases, and private-payment layers. Retroactive disclosure through store-now-decrypt-later and privacy loss (Attack~\ref{atk:retro}) dominates this layer. It can reveal issuer--custodian communications, redemption queues, reserve movements, KYC envelopes, bridge-relayer topology, or shielded stablecoin flows. These failures may not directly mint or steal tokens, but they can expose users, enable targeted compromise, undermine compliance guarantees, and permanently reveal historical payment information.

\pparagraph{Cross-layer implication}
Lower-layer quantum risk propagates upward. A PQ-secure issuer key does not protect users if the host chain's account signatures can be forged via Attack~\ref{atk:on-spend}; a PQ-secure bridge committee does not protect wrapped supply if the source-chain finality evidence can be rewritten via Attack~\ref{atk:at-rest}; and a PQ-secure token contract does not preserve privacy if the surrounding privacy layer is vulnerable to Attack~\ref{atk:retro}, Attack~\ref{atk:equiv}, or Attack~\ref{atk:zk-snd}. Conversely, issuers can still reduce risk before full base-layer migration by hardening custody, admin, bridge, oracle, and off-chain communication paths. The remaining exposure follows the authority--liability gap in Table~\ref{tab:authority-liability}: the actor able to migrate one layer may not control the layer that fails or bear the resulting loss. This layered view motivates the migration strategies in Section~\ref{sec:taxonomy}.

%% file: taxonomy.tex
\section{Taxonomy and Migration Strategies}
\label{sec:taxonomy}

This section provides the main systematization framework of the paper. The \emph{control-surface} dimension identifies who has authority to migrate a component, the \emph{migration-layer} dimension identifies where the change must occur, and the \emph{cryptographic-strategy} dimension identifies how quantum resistance is introduced. Used together, the dimensions expose incomplete migrations: a proposal may introduce a post-quantum primitive while leaving the responsible actor, lower-layer dependency, or legacy state unspecified.

We first summarize the cryptographic dependencies of stablecoins. Table~\ref{tab:crypto_primitives_pq_security} records each primitive's role, representative uses, and current post-quantum direction. Section~\ref{sec:threats} gives the stablecoin consequences of failure.

\begin{table*}[t]
\centering
\scriptsize
\setlength{\tabcolsep}{3.5pt}
\begin{tabularx}{\textwidth}{@{}p{2.1cm}p{3.1cm}p{4.7cm}Y@{}}
\toprule
\textbf{Primitive} & \textbf{Functionality} & \textbf{Blockchain examples} & \textbf{Post-quantum candidates / status} \\
\midrule
Hash functions & Identities, Merkle trees, PoW, commitments, transaction identifiers & SHA-256, RIPEMD-160, Keccak-256, BLAKE2, Merkle trees~\cite{C:Merkle87} & Retain modern hashes with sufficient output length; use 256-bit outputs for long-lived commitments and nullifiers \\
Digital signatures & Account authentication, admin actions, validator messages, contract calls & ECDSA, Schnorr~\cite{JC:Schnorr91}, EdDSA~\cite{JCEng:BDLSY12}, BLS~\cite{AC:BonLynSha01} & ML-DSA and SLH-DSA are standardized~\cite{nistFips204,nistFips205}; FN-DSA/Falcon standardization and nine additional signature candidates remain in progress~\cite{nistPQCProject2026,nistAdditionalSigs2026} \\
Multisignatures & Custody, governance, bridge committees, issuer administration & Schnorr multisig, BLS aggregation, ECDSA multisigs; lattice proposals~\cite{CANS:ElBStu16} & Lattice multisignatures exist; compact aggregation, accountability, key rotation, and wallet integration remain open \\
Threshold signatures & Distributed custody, consensus, bridge attestation, treasury approval & Threshold ECDSA, FROST, threshold BLS & Lattice threshold signatures~\cite{ICALP:AgrSteYad22,AC:ChaTesZhu24}; production performance, identifiable abort, dynamic committees, and custody integration remain open \\
Randomness / VRFs & Leader election, committee sampling, lotteries, validator selection & Ethereum RANDAO~\cite{CCS:NTSL25}; Algorand-style VRFs~\cite{FOCS:MicRabVad99} & Symmetric-key and lattice-based PQ VRFs~\cite{ESORICS:BDEKKL22,EPRINT:Abraham18}; proof size and consensus integration remain limiting \\
Key exchange & Private channels, off-chain coordination, API confidentiality & ECDH, ECIES~\cite{IMA:Smart01} & ML-KEM is standardized; HQC is under standardization as a backup KEM~\cite{nistFips203,nistHQC2025}; hybrid KEMs support staged infrastructure migration \\
Zero-knowledge proofs & Privacy, rollup validity, compliance proofs, reserve proofs & Groth16~\cite{EC:Groth16}, PLONK~\cite{EPRINT:GabWilCio19}, Halo/Halo2~\cite{C:BDFG21,EPRINT:SHJKGA23}, Bulletproofs~\cite{SP:BBBPWM18} & STARKs~\cite{EPRINT:BBHR18}; lattice SNARKs and ZK~\cite{C:ACLMT22,C:LyuNguPla22,C:NguSei22,PKC:GZGPX25}; verifier cost and wallet proving remain open \\
Commitments & Data binding, privacy, confidential balances, reserve proofs & Pedersen~\cite{C:Pedersen91}, EC and KZG polynomial commitments & Hash commitments; lattice commitments~\cite{C:AttLyuSei20}; migration must be coordinated with the proof system \\
Encryption / AEAD & Note encryption, private memos, off-chain communications & ECIES~\cite{IMA:Smart01}, KEM+DEM, symmetric AEAD~\cite{C:BelNgTac19} & ML-KEM or hybrid KEM plus 256-bit AEAD; archived traffic remains subject to SNDL \\
Exotic signatures & Privacy, payment channels, swaps, anonymous authorization & Ring~\cite{AC:RivShaTau01}, blind~\cite{C:Chaum82}, adaptor~\cite{AC:AEEFHM21}, aggregate signatures~\cite{EC:BGLS03} & Lattice ring/blind/adaptor schemes~\cite{ACNS:LuAuZha19,CCS:AKSY22,CCS:BLNS23,ESORICS:EsgErsErk20,FC:TaiMorMaf21}; production readiness and aggregation remain open \\
\bottomrule
\end{tabularx}
\caption{Blockchain cryptographic primitives and post-quantum migration status.}
\label{tab:crypto_primitives_pq_security}
\end{table*}

\pparagraph{Dimension 1: Control surface}
The control-surface dimension asks who has practical authority to change a component. \textbf{Issuer-controlled} components include HSM and MPC policies, smart-contract admin keys, treasury wallets, mint and burn roles, pause and blacklist roles, and emergency multisigs. \textbf{Host-chain-controlled} components include account formats, consensus signatures, base-layer randomness, precompiles, transaction validity, and protocol-level proof systems. \textbf{Ecosystem-controlled} components include custodians, exchanges, oracle providers, bridge operators, relayers, wallet vendors, auditors, and rollup sequencers. A migration plan that names an algorithm but not the actor responsible for deployment is incomplete.

\pparagraph{Dimension 2: Migration layer}
The migration-layer dimension asks where quantum resistance is introduced. \textbf{Infrastructure} changes cover HSMs, MPC platforms, signing ceremonies, encrypted APIs, redemption procedures, and compliance systems. \textbf{Application overlays} include smart-contract wallets, PQ-controlled vaults, hybrid admin gates, claim contracts, and bridge-attestation contracts. \textbf{Bridge/rollup-layer} changes cover lockboxes, cross-chain messages, sequencer authentication, proof systems, withdrawal bridges, and data availability. \textbf{Base-layer} changes cover native account and validator signatures, finality, randomness, light-client verification, gas accounting, and precompiles. Lower-layer risk propagates upward, but issuers can still reduce exposure in higher layers while awaiting consensus changes.

\pparagraph{Dimension 3: Cryptographic strategy}
A \textbf{hybrid} strategy requires both classical and PQ authorization or binds a classical key to a PQ key during transition. A \textbf{PQ-native} strategy removes the classical key from the relevant authorization path. \textbf{Encapsulation} places an existing asset, key, or contract behind a PQ-aware vault or wrapper. Encapsulation is deployable without changing the original token, but remains dependent on the original admin key, host chain, bridge lockbox, or finality assumptions. It is therefore a partial migration rather than an end state. A related class of measures reduces the exposure window of a classical key without introducing post-quantum cryptography: hash-and-rotate accounts that commit only to a key hash and reveal each key once, and encrypted mempools that hide a transaction until ordering~\cite{circlePQRoadmap2026}. These shorten the on-spend race (Attack~\ref{atk:on-spend}) but do not protect at-rest keys, so we treat them as complements to a hybrid strategy rather than as a fourth strategy.

\pparagraph{Applying the taxonomy}
Issuer-controlled migration can begin with mint, burn, treasury, reserve, oracle, redemption, and emergency-administration procedures. HSM/MPC approvals, signing ceremonies, APIs, and node links can use PQ or hybrid authentication before base-layer account formats change. This does not make the final classical on-chain signature PQ-secure, but it reduces infrastructure exposure and tests key lifecycle, audit, and recovery. Where verification is available, issuers can place privileged controls in hybrid or PQ-aware contract wallets and vaults. Crypto-collateralized systems should prioritize oracle, governance, liquidation, and emergency-shutdown paths; bridged systems should prioritize attestation, source-finality, lockbox, and redemption controls.

\pparagraph{Dormant, rollup, and wrapped state}
A migration must state what happens to users who do nothing. Options include voluntary migration with incentives, redemption and reissuance, claim contracts, a freeze after a deadline, or a post-quantum recovery proof. An issuer has additional levers that a base-layer protocol lacks: it can pause contract functionality on, or withdraw support from, a host chain that misses a post-quantum deadline, and it can relocate a copy of a frozen contract to a chain it controls for managed recovery~\cite{circlePQRoadmap2026}. Each option changes property and governance semantics. Rollups require separate treatment of sequencer keys, proof systems, data-availability committees, settlement contracts, and withdrawal bridges. Wrapped supply requires route-specific sunset dates, redemption windows, and rules for residual liquidity. A PQ-safe L1 account does not automatically protect L2 or wrapped balances.

\subsection{Current Proposals and Deployment Directions}
\label{sec:proposals}

Table~\ref{tab:proposals} applies the dimensions to recent proposals. These are drafts, roadmaps, or product-specific systems unless stated otherwise. They demonstrate migration mechanisms, but should not be read as deployed guarantees.

\begin{table*}[t]
\centering
\scriptsize
\setlength{\tabcolsep}{3.5pt}
\begin{tabularx}{\textwidth}{@{}p{2.15cm}p{2.25cm}p{2.7cm}Y p{3.0cm}@{}}
\toprule
\textbf{Proposal/system} & \textbf{Control/layer} & \textbf{Mechanism} & \textbf{What it addresses} & \textbf{Stablecoin limitation} \\
\midrule
Zcash ZIP-2005~\cite{zcashzip2005} & Host chain; privacy & Pre-crisis recoverability hooks for Orchard notes & Preserves a future recovery route after discrete-log assumptions fail & Does not specify the final PQ proof system, issuer policy, or redemption treatment \\
Bitcoin BIP-360/361~\cite{bip360,bip361} & Host chain; outputs and governance & P2MR output plus phased legacy-signature sunset and possible recovery & Reduces exposure for new outputs and makes dormant-state policy explicit & Does not migrate stablecoin contracts or resolve coercive freeze/recovery semantics \\
Ethereum EIP-8051~\cite{eip8051} & Base layer; precompile & ML-DSA and EVM-optimized verification precompiles & Proposes 4,500-gas specialized verification & Draft gas model; expanded key representation; no account or contract migration policy \\
Ethereum EIP-8164~\cite{eip8164} & Base layer; account & Permanent ECDSA-to-ML-DSA-44 native key delegation & In-place account migration and PQ key rotation; proposes 50,000 intrinsic gas & Draft; does not migrate contracts, bridges, validators, or dormant users \\
Ethereum PQ roadmap~\cite{ethereumPQ2026} & Execution, consensus, data & PQ verification, account agility, hash-based validators, proof aggregation & Treats migration as a multi-layer program rather than one signature change & Timelines, validator aggregation, and production mechanisms remain research targets \\
Algorand State Proofs / PQ roadmap~\cite{algorandPQRoadmap2026,SP:MRVWZ21} & History, accounts, institutional control, consensus & Falcon-signed State Proofs and AVM verification; planned native/hybrid PQ accounts and multisig & Combines deployed historical attestations with an explicit path for treasury and custody accounts & PQ VRF and consensus signatures remain research/deployment targets; roadmap dates are forward-looking \\
Starknet PQ roadmap~\cite{starknetPQRoadmap2026,EPRINT:BBHR18} & Rollup, accounts, state commitments & STARK proving, native account abstraction, Pedersen-to-BLAKE2 migration, early Falcon-512 accounts & Separates proof-layer and account-layer migration; account verifiers can change without a protocol fork & Early PQ accounts are pre-audit; bridge syscalls and EIP-4844 KZG inherit Ethereum dependencies \\
EdDSA-chain readiness~\cite{EPRINT:BalChaRoy25} & Host chain; account & Registration and transition using existing EdDSA account structure & Preserves deployed address formats while adding PQ credentials & Chain-specific; application, issuer, and bridge controls still migrate separately \\
Circle xReserve/CCTP~\cite{circleXReserve2025,circleCCTPDeprecation2025} & Issuer and partner chains; bridge & Issuer attestations, reserve contract, partner-chain mint/burn logic, version deprecation & Shows an issuer-controlled cross-chain migration and legacy-version sunset surface & Attestation API, contracts, partner chains, liquidity, and legacy routes must transition coherently \\
Circle PQ roadmap~\cite{circlePQRoadmap2026} & Issuer; infrastructure, application, and (on Arc) base layer & SLH-DSA-SHA2-128s precompile, dual-mode USDC contracts with a classical-disable switch, multisig cold storage, TEE privacy layer, phased switch with recovery policy & The only issuer-authored plan; states which surfaces the issuer controls and which it defers to host chains, HSM and cloud vendors, wallets, and contract owners & Base-layer parts apply to Arc only; PQ consensus signature unselected pending aggregation; PQ threshold custody deferred; enclave attestation remains classical \\
\bottomrule
\end{tabularx}
\caption{Recent proposals address different parts of the post-quantum migration problem.}
\label{tab:proposals}
\end{table*}

The Algorand and Starknet rows summarize their June 2026 technical roadmaps~\cite{algorandPQRoadmap2026,starknetPQRoadmap2026}. The Circle row is the only issuer-authored plan in the table and is useful precisely because it separates the three control surfaces of Dimension~1 from one actor's viewpoint. As issuer, Circle plans dual-mode contracts, multisig cold storage, and infrastructure key rotation; as operator of the Arc chain, it plans a signature precompile, an \texttt{ecrecover} replacement, and validator-signature upgrades; and it explicitly defers HSM post-quantum support, enclave attestation, wallet key derivation, remediation of immutable contracts, and the migration of more than thirty other host chains to third parties~\cite{circlePQRoadmap2026}. The deferred items are the authority--liability gap stated by a party on the authority side. The Ethereum proposals also show why gas figures require an execution model. EIP-8051 assumes a specialized precompile and expanded verification key, whereas EIP-8164 assigns an intrinsic protocol cost to native account authentication. Neither is the cost of verifying ML-DSA in ordinary Solidity. By contrast, \emph{poqeth} measures full EVM verification and finds costs from hundreds of thousands to hundreds of millions of gas for the studied schemes and parameters~\cite{ASIACCS:KSKK25}. Appendix~\ref{app:costs} separates measured, proposed, and calculated figures.

\pparagraph{Operational sequence}
A practical stablecoin migration can be staged without treating the stages as independent. First, inventory assumptions by monetary function---mint, burn, treasury, upgrade, pause, blacklist, redemption, oracle, governance, bridge, rollup, privacy, and settlement---and record the controlling actor and loss bearer. Second, harden issuer-controlled infrastructure and test key rotation and recovery. Third, add application, bridge, and oracle safeguards such as hybrid gates, route limits, versioned attestations, and authenticated checkpoints. Fourth, use host-chain signature agility or precompiles to remove classical authentication. Finally, define the checkpoint, claim, redemption, deadline, and loss-allocation rules for dormant and wrapped state. Auditors should retain key provenance, signer rosters, dependency graphs, test vectors, recovery exercises, and evidence of redemption continuity.

\subsection{Representative Stablecoin Designs and Deployments}

Table~\ref{tab:stablecoin-taxonomy} maps representative stablecoin classes to the dependencies developed above. It is not a fourth taxonomy dimension; it shows how the same dimensions specialize across stablecoin designs.

\begin{table*}[t]
\centering
\scriptsize
\setlength{\tabcolsep}{3.5pt}
\begin{tabularx}{\textwidth}{@{}p{3.1cm}p{3.0cm}p{3.2cm}Y@{}}
\toprule
\textbf{Stablecoins} & \textbf{Design} & \textbf{Deployment} & \textbf{Cryptographic dependencies} \\
\midrule
USDT, USDC, PYUSD, EURC & Fiat-backed, centralized issuer & Ethereum, Tron, Solana, Base, and other chains & Account signatures, issuer roles, custody and treasury keys, upgrade controls, native or cross-chain transfer systems \\
DAI & Crypto-collateralized & Ethereum and L2 deployments & Accounts, governance, oracle signatures, liquidation logic, collateral contracts, keeper infrastructure \\
LUSD & Crypto-collateralized, low-governance design & Ethereum & Accounts, immutable or low-governance contracts, liquidation logic, price feeds \\
FRAX and related hybrid designs & Hybrid collateral and algorithmic mechanisms & Ethereum and L2 deployments & Governance keys, oracle feeds, AMM/liquidity logic, collateral accounting, account signatures \\
UST and earlier algorithmic designs & Algorithmic/seigniorage-style & Historical deployments & Governance, oracle or market mechanisms, account signatures, liquidity and arbitrage infrastructure \\
USDC.e and wrapped USDC/USDT variants & Bridged or wrapped representation & Bridges, rollups, sidechains, app-chains & Attestations, lockboxes, wrapped-token contracts, light clients, relayers, committee or validator signatures \\
Stablecoins in Aztec- or Railgun-style private systems; Circle Research--Inco design & Private or shielded transfer layer & Shielded pools, private rollups, confidential-token systems & Commitments, encryption or FHE, nullifiers, ZK proofs, viewing/disclosure mechanisms, underlying token controls~\cite{circleInco2024} \\
\bottomrule
\end{tabularx}
\caption{Representative stablecoin classes and their main cryptographic dependencies.}
\label{tab:stablecoin-taxonomy}
\end{table*}

The table highlights two recurring patterns. First, stablecoins inherit lower-layer cryptography even when their own contracts are simple token contracts. Second, stablecoin-specific risk is concentrated in mint keys, treasury accounts, bridge attestations, oracles, custody procedures, and upgrade paths that can affect much more value than an ordinary user key.

\pparagraph{Cost evidence}
The evidence in Appendix~\ref{app:costs} supports three conclusions. Native verification latency may be acceptable for low-frequency issuer and custody actions, but calldata, data availability, aggregation, and verifier support dominate bridge, oracle, rollup, and user-transfer costs. Available measurements permit size and some verifier-cost estimates, but do not evaluate complete stablecoin operations such as mint authorization, threshold custody withdrawal, bridge redemption, oracle retry, dormant-balance claims, private transfers, or rollup settlement. Section~\ref{sec:research-gaps} treats this missing evidence as a research problem rather than presenting microbenchmarks as an end-to-end evaluation.

%% file: gaps.tex
\section{Research Insights, Gaps, and Open Problems}
\label{sec:research-gaps}

We now organize the research agenda around four stablecoin-specific findings and develop the corresponding open problems, as the preceding sections show that post-quantum stablecoin security is not obtained by substituting one signature scheme for another.

\subsection{Standardized Primitives Do Not Cover Stablecoin Functions}
\label{sec:functionality-gap}

\pparagraph{Finding}
NIST-standardized signatures and key encapsulation provide necessary building blocks for issuer administration, wallet authentication, and encrypted issuer--custodian channels. They do not cover the full functions that stablecoin systems require. The relevant unit of migration is therefore a \emph{stablecoin function}, not an isolated primitive.

For issuer and custody keys, a replacement must support threshold policy, recovery, audit, HSM/MPC integration, and signer rotation. For bridges and oracles, it must support compact, accountable committee authorization at high message frequency. For private or compliance-enabled transfers, it must replace vulnerable commitments, proofs, encryption, and disclosure mechanisms without making mobile proving or on-chain verification impractical. For host-chain consensus, the loss of BLS aggregation becomes a bandwidth and finality problem rather than merely a larger-signature problem. Ethereum's public post-quantum program, for example, considers hash-based validator signatures together with proof aggregation~\cite{ethereumPQ2026}.

\pparagraph{Gap: authorization suites rather than isolated schemes}
Post-quantum multisignatures, threshold signatures, VRFs, adaptor signatures, and proof systems exist, but they have not yet been combined into production authorization suites for stablecoin operations. Recent threshold ML-DSA constructions based on MPC and identifiable-abort techniques narrow one part of this gap~\cite{EPRINT:BCEPT25,EPRINT:BCdENP25}, but they do not by themselves provide the complete custody and governance semantics required by a stablecoin deployment. A useful design must specify committee churn, identifiable abort, key recovery, audit logs, hardware assumptions, signature or proof size, verification location, and failure handling. The requirements differ across four workloads: (i) issuer and custody control, (ii) bridge and oracle committees, (iii) private and compliance-enabled transfers, and (iv) host-chain or rollup settlement. Security proofs are necessary, but insufficient if a scheme cannot reproduce the control and audit semantics of the system it replaces. Deployment evidence confirms the gap. Circle's roadmap judges post-quantum threshold and MPC signing prohibitive in the near term across the algorithms it must support and instead moves cold storage to on-chain multisig contracts; it also warns that exporting keys from an HSM without post-quantum support in order to sign on a general-purpose CPU creates a larger classical risk than the quantum risk it is meant to address~\cite{circlePQRoadmap2026}.

\pparagraph{Gap: implementation and custody security}
Algorithmic post-quantum security does not imply secure deployment in custody hardware. Practical side-channel work on Dilithium/ML-DSA and Falcon shows that lattice signatures introduce implementation surfaces that require dedicated hardening~\cite{PQCRYPTO:SLKG23,EPRINT:KarAys21}. For stablecoin issuers and custodians, evaluation should therefore include secure-element and HSM memory limits, key-generation and signing latency, masking and fault protections, key-export requirements, and compatibility with threshold or MPC custody, rather than only signature size and verifier cost.

\pparagraph{Gap: aggregation and economic amplification}
Standardized PQ signatures do not provide BLS-like non-interactive aggregation. Posting independent signatures increases calldata, data availability, and verifier work, and the increase is amplified by bridge messages, oracle updates, validator votes, and rollup settlement. Research should compare threshold signing, sequential or interactive aggregation, proof aggregation, committee sampling, and optimistic verification under the same accountability and failure model. The objective is not simply the smallest signature; it is the lowest end-to-end cost that preserves attribution, slashing or dispute evidence, and recovery. The same constraint appears in issuer planning: Circle leaves its post-quantum consensus signature unselected pending a practical aggregate alternative to BLS, and considers STARK batch verification and fraud proofs as throughput options~\cite{circlePQRoadmap2026}. The fraud-proof route rests on signatures that decompose into independently checkable components, so that a validator who attests to a batch can be refuted by exhibiting one failing component; flexible signatures, which allow verification to be truncated at a chosen cost and still yield a graded validity guarantee, provide one such structure~\cite{ESORICS:LeKelKat19}.

\subsection{Security Level and Cost Depend on the Stablecoin Operation}
\label{sec:security-level}

\pparagraph{Finding}
Grover's quadratic speedup does not make every hash or symmetric component equally urgent. With well-sized 256-bit primitives and poor linear parallelization, public-key authentication and loss of compact aggregation dominate the immediate migration problem. However, the appropriate post-quantum security level is operation-specific. A treasury recovery key may be used rarely but protect enormous value for decades. An oracle message may carry less value per update but be issued continuously. A retail transfer must fit wallet and fee constraints. A bridge attestation combines high value and high frequency.

\pparagraph{Representative cost anchors}
Table~\ref{tab:representative-benchmarks} puts the workload
differences above against representative cryptographic costs.
The numbers should not be compared as if they measured the same
execution model: signature sizes come from standards or scheme
specifications, calldata values account only for transporting the
signature, and protocol-supported verification can differ by orders
of magnitude from ordinary smart-contract verification. The table
is therefore useful for identifying the dominant constraint for each
stablecoin operation, rather than for ranking schemes by a single
cost metric.

\begin{table*}[t]
\centering
\scriptsize
\setlength{\tabcolsep}{3.5pt}
\begin{tabularx}{\textwidth}{
  @{}p{2.15cm}p{2.25cm}p{2.65cm}p{3.15cm}Y@{}}
\toprule
\textbf{Mechanism}
& \textbf{Key / signature size}
& \textbf{Native or protocol support}
& \textbf{Data / on-chain implication}
& \textbf{Stablecoin interpretation} \\
\midrule

ECDSA / secp256k1
& 65-B signature
& Native EVM \texttt{ecrecover}
& 3,000 gas verification; about 1,040 calldata gas
  \cite{ethereumYellowPaper,eip2028Calldata}
& Current account and administrator baseline. \\

ML-DSA-44
& pk 1,312 B; sig 2,420 B \cite{nistFips204}
& Standardized; proposed specialized Ethereum support in
  EIP-8051 and EIP-8164
& About 38.7k calldata gas for the signature alone;
  verification depends strongly on execution support
  \cite{eip2028Calldata,eip8051,eip8164}
& Plausible for issuer administration, mint/burn, and
  low-frequency treasury authorization. \\

ML-DSA-65
& pk 1,952 B; sig 3,309 B \cite{nistFips204}
& Standardized; no deployed EVM precompile
& About 52.9k calldata gas for the signature alone
  \cite{eip2028Calldata}
& Illustrates the bandwidth cost of higher-security
  bridge, oracle, or custody authorization. \\

Falcon-512
& pk 897 B; sig about 666 B \cite{NISTPQC:FALCON22}
& Compact lattice signature; FN-DSA standardization is
  still in progress
& About 10.7k calldata gas for a typical signature;
  no deployed EVM precompile
& Attractive when bandwidth dominates, but implementation
  and side-channel requirements complicate custody/HSM use. \\

SLH-DSA
& Parameter-dependent; signatures are several to tens of KB
  \cite{nistFips205}
& Standardized hash-based construction
& An 8-KB signature alone costs over 100k calldata gas;
  contract verification is substantially more expensive
& Better suited to infrequent recovery or emergency paths
  than high-frequency bridge or oracle messages. \\

PQ-friendly proofs
& STARK and lattice-proof sizes are workload-dependent
& Hash- and lattice-based constructions avoid pairing/DLP
  assumptions
& No meaningful single gas figure: cost depends on
  arithmetization, recursion, aggregation, and verifier design
& Relevant to private transfers, reserve/compliance proofs,
  and rollup settlement. \\

\bottomrule
\end{tabularx}
\caption{Representative cost anchors for stablecoin-relevant
classical and post-quantum primitives. Size-derived calldata costs
use 16 gas per non-zero byte and exclude verification, public-key
storage, aggregation, and transaction overhead. Detailed measured
and proposed EVM costs appear in Appendix~\ref{app:costs}.}
\label{tab:representative-benchmarks}
\end{table*}

\begin{table}[t]
\centering
\scriptsize
\setlength{\tabcolsep}{3pt}
\begin{tabularx}{\columnwidth}{@{}p{1.45cm}p{1.45cm}p{1.45cm}Y@{}}
\toprule
\textbf{Operation} & \textbf{Frequency} & \textbf{Value concentration} & \textbf{Primary constraint} \\
\midrule
Treasury or recovery & Low & Very high & Long-term assurance, audit, recovery \\
Mint, burn, admin & Low--medium & Very high & Direct verification, control continuity \\
Bridge/oracle & High & High & Aggregation, bandwidth, latency \\
Retail transfer & Very high & Low per transfer & Fee, wallet size, batching \\
Private transfer & High & Variable & Proof size, proving time, disclosure policy \\
\bottomrule
\end{tabularx}
\caption{Post-quantum requirements differ across stablecoin operations.}
\label{tab:workload}
\end{table}

\pparagraph{Gap: choosing a security level}
Standards bodies classify security levels against defined cryptanalytic benchmarks, whereas blockchain deployments often make cost-driven choices based on gas, bytes, bandwidth, wallet support, and validator throughput. The lowest standardized level may still be costly on-chain, creating pressure to use weaker or ad hoc constructions. Stablecoin research should select parameters from an explicit risk budget that includes value at risk, key lifetime, public-key exposure, migration latency, update frequency, and the consequences of a failed authorization. The result may legitimately differ between a cold treasury key and a high-frequency oracle, but the difference should be justified rather than inherited from unrelated conventions. Circle's roadmap applies this reasoning explicitly: it selects level-1 SLH-DSA-SHA2-128s for signatures, which can be upgraded later, but level-3 hybrid KEMs (X-Wing, X25519MLKEM768) for encryption, because privacy lost to store-now-decrypt-later cannot be restored~\cite{circlePQRoadmap2026}.

\pparagraph{Gap: complete-operation benchmarks}
Table~\ref{tab:representative-benchmarks} provides useful
primitive-level cost anchors, but these measurements are not sufficient
to optimize complete stablecoin operations. \emph{poqeth} shows that
EVM opcode pricing and verifier design can dominate native cryptographic
speed~\cite{ASIACCS:KSKK25};
EIP-8051 and EIP-8164 propose very different costs because they assume different protocol support and key representations~\cite{eip8051,eip8164}. Calldata makes a 2,420-byte ML-DSA-44 signature roughly 37 times larger than a 65-byte ECDSA signature, but that comparison omits aggregation, public-key storage, contract execution, and failure recovery. A complementary deployment example comes from Liquid: Blockstream deployed a SHRINCS verifier as a Simplicity spending condition on Liquid mainnet, allowing opt-in PQ protection for issued assets including stablecoins without a consensus change; transaction size is tied to the verification budget, while the peg, Confidential Assets commitments, and block-signing consensus remain classical~\cite{blockstreamLiquidPQ2026}. This illustrates why verifier placement and residual dependencies belong in an operation-level benchmark.

A stablecoin benchmark suite should therefore measure complete operations: PQ mint authorization, threshold custody withdrawal, bridge attestation and redemption, oracle update and retry, user or dormant-balance migration claims, private transfer and selective disclosure, and rollup settlement. It should report gas or native verifier cost, bytes, data-availability cost, latency, signer throughput, wallet memory, HSM/MPC support, audit evidence, recovery behavior, and financial value exposed during the operation. Appendix~\ref{app:costs} summarizes the current evidence and its limitations.

\subsection{The Authority--Liability Gap Creates Residual and Legacy Risk}
\label{sec:authority-gap}

\pparagraph{Finding}
Table~\ref{tab:authority-liability} shows that each major stablecoin control surface crosses an organizational boundary. An issuer may rotate mint keys but cannot update user accounts; a host chain may add a new account type but cannot force a bridge or oracle to adopt it; a bridge may migrate its committee while exchanges continue to trade residual wrapped supply. The actor able to mitigate the risk is often not the party that bears the loss, which creates strategic delay and incomplete protection.

\pparagraph{Gap: cross-layer residual risk}
A stablecoin can harden issuer keys while remaining exposed through accounts, consensus, a bridge, an oracle, a rollup, a custodian, or a privacy layer. Conversely, a host chain can add PQ accounts while a stablecoin remains vulnerable through upgrade keys, minting authority, bridge attestations, or proof systems. Formal models should represent residual assumptions across layers rather than label one component ``PQ-safe.'' A useful model must identify the controlling actor, inherited assumptions, value exposed, and state transition after another layer fails.

\pparagraph{Gap: explicit authority and liability rules}
Migration protocols should specify who declares an emergency, which checkpoint is authoritative, how users prove claims, how native and wrapped supply are reconciled, what happens after a deadline, how appeals and recovery work, and who absorbs losses from a late or failed migration. This analysis must combine cryptographic recovery with property, redemption, accounting, and liquidity consequences. Treating all participants as a single coordinated upgrader misses the central stablecoin problem.

\pparagraph{Gap: bridge and wrapped-supply asymmetry}
A destination representation depends on the source lockbox, attestation service, source finality, relayers, destination contract, exchanges, and liquidity venues. These components can migrate at different times. A technically secure new route does not extinguish claims or liquidity created by an old route. Research is needed on route versioning, authenticated checkpoints, redemption windows, legacy-route shutdown, detection of unbacked supply, and treatment of residual tokens after the old assumptions become unsafe.

\pparagraph{Gap: dormant balances and coercive migration.}
Leaving dormant balances untouched preserves current ownership rules but leaves
a persistent quantum target. Freezing or recovering them reduces theft risk but
changes control and may conflict with redemption, due-process, or
customer-protection duties. The same governance dilemma has been identified
for unmigrated cryptocurrency balances, with proposals ranging from leaving
vulnerable assets spendable to freezing them after a deadline or introducing
special recovery mechanisms~\cite{coinbaseAbandonedCoins2026}.
Circle's post-quantum roadmap gives the issuer's side of this
dilemma~\cite{circlePQRoadmap2026}. It separates disabling unsafe cryptographic
control from extinguishing an asset holder's economic interest, treats a
post-cutoff freeze as a protective control rather than a finding of
abandonment, and lists recovery routes: relocating frozen contracts to an
issuer-controlled chain, zero-knowledge seed-recovery proofs for BIP-32 and
SLIP-0010 hierarchies, TEE-attested seed verification, and off-chain evidence
of entitlement. It identifies escheat and bailment as the closest legal
analogues and asks regulators for guidance on notice periods, evidentiary
standards, and recoverability windows. This contrasts with BIP-361's
rescue-protocol encumbrance~\cite{bip361} and Optimism's announced January 2036
deadline as reported in the roadmap, neither of which specifies a recovery route.
Stablecoin issuers thus have more administrative tools than base-layer protocols,
but those tools make the governance choice more explicit rather than eliminating
it. Migration mechanisms should state the trigger, notice period, proof of
ownership, deadline, recovery route, and effect on redemption for users who do
not or cannot act.

\subsection{Privacy, Measurement, and Regulation Must Become Testable}
\label{sec:measurement-policy}

\pparagraph{Finding: private stablecoins use different cryptographic stacks}
Private stablecoin designs do not share a single cryptographic stack.
The Circle Research--Inco confidential ERC-20 design uses fully
homomorphic encryption over encrypted balances and transfer amounts,
whereas note-based systems usually combine commitments,
zero-knowledge proofs, nullifiers, viewing keys, and encrypted note
data~\cite{circleInco2024}. A third stack, planned for Circle's Arc chain,
executes standard contracts inside trusted execution environments with
X-Wing HPKE transaction encryption and symmetric state encryption; its
post-quantum privacy claim rests on enclave attestation and certificates
that remain classical until the cloud provider upgrades
them~\cite{circlePQRoadmap2026}. Each stack fails differently under a
CRQC, and the residual dependency of the TEE design lies with a vendor
outside the issuer's control.

\pparagraph{Gap: quantum-vulnerable supply measurement}
Market capitalization alone obscures the distribution of risk among exposed user keys, issuer roles, bridge lockboxes, oracle committees, rollup proof systems, and private pools. We propose reporting \emph{quantum-vulnerable supply} as a vector rather than one scalar:
\[
\mathrm{QVS}=\bigl(Q_{\mathrm{user}},Q_{\mathrm{issuer}},Q_{\mathrm{bridge}},Q_{\mathrm{oracle}},Q_{\mathrm{L2}},Q_{\mathrm{privacy}}\bigr),
\]
where each component measures stablecoin value whose continued integrity depends on a specified quantum-vulnerable assumption. Components should be reported per chain and representation, with explicit treatment of shared keys and overlapping dependencies to avoid double counting. Public tools should map contract roles, key exposure, bridge routes, oracle signer sets, rollup and proof dependencies, dormant balances, and migration status.

\pparagraph{Gap: governance and regulatory readiness}
Regulation increasingly requires reserve control, redemption, operational resilience, governance, and intervention capabilities, but post-quantum readiness is rarely expressed as a testable control. MiCA recovery and redemption duties imply that an issuer should demonstrate continued redemption under key loss, contract migration, or a freeze-and-reissue event~\cite{mica2023}. The GENIUS Act and OCC proposals imply tests for preserving mint, burn, freeze, transfer-prevention, custody, screening, blocking, and reporting capabilities after a cryptographic transition~\cite{geniusActText2025,occGenius2026,occGeniusAml2026}. Audits should therefore report both technical readiness and continuity of redemption, reserve control, legally required intervention, and historical evidence after classical signatures lose evidentiary value.

\pparagraph{Research agenda}
The strongest open problems are compositional. Future work should build stablecoin-specific benchmark suites, formal models of residual risk and authority, migration protocols for dormant and wrapped supply, efficient aggregate and threshold attestations for bridges and custody, PQ-friendly private and compliance-enabled transfers, and longitudinal measurements of quantum-vulnerable supply. The objective is an end-to-end transition that preserves authorization, redemption, supply integrity, finality, privacy, and required controls across every layer.

%% file: conclusion.tex
\section{Conclusion}

Stablecoins sit at the intersection of public blockchains, traditional finance, cross-chain infrastructure, and regulation. Their concentrated issuer, custody, oracle, bridge, and upgrade controls make them unusually sensitive to a CRQC, while their host-chain, rollup, and privacy dependencies prevent any one actor from completing the migration alone. The central stablecoin-specific problem is the authority--liability gap: the party able to change a component is often different from the users, exchanges, protocols, or issuers that bear the loss if it fails.

A practical transition should begin with a function-level cryptographic inventory, protect issuer-controlled infrastructure, add contract, bridge, and oracle safeguards, adopt host-chain signature agility, and define explicit treatment of dormant and wrapped supply. Standardized KEMs and signatures provide necessary building blocks, but do not by themselves replace aggregate bridge attestations, threshold custody, private and compliance-enabled transfers, rollup proofs, or recovery governance. Regulation further constrains the result by requiring continuity of redemption, reserve control, audit evidence, and intervention capabilities.

The research agenda is therefore compositional. It includes complete-operation benchmarks, formal models of layered residual risk and migration authority, efficient aggregate and threshold authorization, post-quantum privacy and compliance mechanisms, protocols for dormant balances and legacy wrapped routes, and empirical measurement of quantum-vulnerable stablecoin supply. Progress should be measured not by whether one key has been replaced, but by whether authorization, redemption, supply integrity, finality, privacy, and legally required controls remain valid across the full stablecoin system.

%% file: appendix.tex
\section{Supplementary Cryptographic Details}
\label{app:definitions}

This appendix provides formal  definitions for Section~\ref{sec:background}. 

\pparagraph{Hash functions}
A hash function family $\hashf : \mathcal{K} \times \mathcal{M} \to \bset^{\lambda}$
maps an arbitrary-length message to a fixed-length digest.
 
\begin{property}[Hash security]
\label{prop:hash}
A hash function should satisfy \textbf{preimage resistance} (given $y = \hashf(m)$,
find $m'$ with $\hashf(m') = y$), \textbf{second-preimage resistance} (given $m$,
find $m' \neq m$ with $\hashf(m') = \hashf(m)$), and \textbf{collision resistance}
(find any $m \neq m'$ with $\hashf(m) = \hashf(m')$).
\end{property}
 
\noindent \emph{Quantum impact.} Grover~\cite{STOC:Grover96} gives a quadratic preimage
speedup ($\Theta(2^{\lambda/2})$), and BHT \cite{brassard1997quantum} gives $\Theta(2^{\lambda/3})$ for
collisions assuming quantum RAM; under the realistic parallel/low-memory model
the classical $2^{\lambda/2}$ collision bound dominates. Grover parallelizes
poorly (a $\sqrt{k}$ speedup over $k$ instances), so 256-bit hashes (SHA-256,
SHA3-256, Keccak-256, BLAKE2b/3) retain $\approx 128$-bit post-quantum security.
 
\noindent \emph{Post-quantum stance.} No change of primitive is required; use 256-bit
outputs everywhere Merkle roots, nullifiers, or commitments must remain
unforgeable. Hash-based signatures (\SLHDSA, XMSS, LMS) and STARK proof
systems~\cite{NISTPQC:SPHINCS+22,PQCRYPTO:BucDahHul11,EPRINT:KamFlu17,EPRINT:BBHR18} inherit post-quantum (PQ) security entirely from these properties. 
 
\pparagraph{Digital signatures}
A signature scheme $\Sigma = (\KeyGen, \Sign, \Verify)$ is required to satisfy
\emph{existential unforgeability under chosen-message attack} (EUF-CMA): no
$\PPT$ adversary with signing-oracle access can produce a valid signature on a
fresh message except with negligible probability.
 
\noindent \emph{Quantum impact.} Every blockchain-deployed scheme based on DLP/IFP
(ECDSA, Schnorr~\cite{JC:Schnorr91}, EdDSA~\cite{JCEng:BDLSY12},
RSA-PSS, BLS~\cite{AC:BonLynSha01}) falls to
Shor~\cite{FOCS:Shor94}: $\sk$ is recovered from $\pk$ without any oracle access,
so EUF-CMA is broken outright. Crucially, \emph{non-repudiation is broken
retroactively}; a forgery produced after a CRQC is cryptographically
indistinguishable from an honest historical signature, undermining audit logs,
reserve attestations, and legal evidence. The Google Quantum AI estimate
\cite{babbush2026prx} distinguishes the \emph{at-rest} regime ($\pk$ already on
chain) from the \emph{on-spend} regime ($\pk$ revealed at spending time);
issuer, treasury, bridge, and oracle keys are almost always at-rest.
 
\noindent \emph{Post-quantum candidates.} NIST has standardized ML-DSA and SLH-DSA and is continuing
standardization of FN-DSA, based on Falcon:
\MLDSA (FIPS~204~\cite{nistFips204}, derived from
\Dilithium~\cite{NISTPQC:CRYSTALS-DILITHIUM22}), lattice-based, $\approx$1.3--2.6\,KB
public keys and 2.4--4.6\,KB signatures, the default for issuer admin and
oracle workloads; \SLHDSA (FIPS~205~\cite{nistFips205}, derived from
\SPHINCS), stateless hash-based with 8--49\,KB signatures, attractive for
cold-path treasury actions; and \Falcon~\cite{NISTPQC:FALCON22}, NTRU lattice,
compact ($\approx$666\,B) but with side-channel-sensitive Gaussian sampling.
Code- (CROSS, LESS), multivariate- (UOV, MAYO), and isogeny-based (SQIsign)
families are less mature. Critically, \emph{none support non-interactive aggregation} comparable to BLS, creating an important limitation for bridge committees.
 
\pparagraph{Multi-signatures and threshold signatures}
A $(t,n)$-threshold signature scheme distributes signing across $n$ parties so
that any $\geq t$ can jointly sign and fewer than $t$ corruptions cannot, with
additional desiderata of \emph{robustness} and \emph{identifiable abort}.
Threshold ECDSA, threshold Schnorr (FROST), and threshold BLS~\cite{CCS:GenGol18,SAC:KomGol20,CCS:BacLos22,AC:BonLynSha01} all reduce to
DLP/pairing assumptions; under Shor, the aggregate $\sk$ is recovered directly
from the aggregate $\pk$, so the threshold policy provides no protection
whatsoever: a quantum-empowered single adversary acts as if all $n$ shares
were compromised. This is severe because most institutional custody, exchange
hot wallets, and bridge committees deploy a threshold or multi-sig over a
DL-based scheme. Lattice-based multi- and threshold-signature constructions
exist~\cite{CANS:ElBStu16,ICALP:AgrSteYad22,AC:ChaTesZhu24}, but production
gaps remain in dynamic committees, identifiable abort, HSM/MPC integration,
and aggregate sizes within current calldata budgets.
 
\pparagraph{Key exchange and authenticated channels}
A KEM $(\KeyGen, \Encaps, 
\allowbreak \Decaps)$ is \emph{IND-CCA secure} if no $\PPT$
adversary can distinguish a real encapsulated key from uniform even with
decryption oracle access. Classical DH/ECDH, ECIES~\cite{IMA:Smart01}, and
RSA-OAEP all reduce to DLP or factoring and lose IND-CCA security under Shor.
The dominant pattern is \emph{store-now-decrypt-later} (SNDL): adversaries
archive encrypted issuer--custodian traffic today and decrypt it once a CRQC
exists, exposing signing-ceremony schedules, reserve instructions, compliance
correspondence, and bridge relayer topology. ML-KEM is standardized in FIPS~203~\cite{nistFips203};
HQC~\cite{NISTPQC-R4:HQC22} has been selected for standardization
as a backup KEM.
Production
deployments use \emph{hybrid KEMs} (e.g., X25519+\MLKEM-768 in TLS~1.3) that
remain secure as long as either component does, giving a clean transition
story for issuer APIs and custodian links. 
 
\pparagraph{Randomness, VRFs, and leader election}
A verifiable random function (VRF) is $(\KeyGen, \Eval, \Verify)$ where
$\Eval_{\sk}(x) \to (y, \pi)$.
 
\begin{property}[VRF security]
\label{prop:vrf}
A VRF satisfies \textbf{uniqueness} (at most one $y$ verifies for given $\pk, x$),
\textbf{pseudorandomness} ($y$ is indistinguishable from uniform without $\sk$,
even given proofs on other inputs), and \textbf{public verifiability}.
\end{property}
 
\noindent \emph{Quantum impact.} Deployed EC-VRFs (Algorand, Cardano/Chainlink) base both
uniqueness and pseudorandomness on DLP; under Shor, $\sk$ is recovered,
\textbf{pseudorandomness is fully lost} (all future outputs become predictable),
and \textbf{uniqueness may also fail} depending on construction. Threshold-BLS
beacons (drand) are broken at the threshold level, allowing biased or
fully-controlled beacon outputs. For stablecoins this affects PoS
validator/committee selection, liquidation auctions, and any governance lottery.
 
\smallskip

\noindent \emph{Post-quantum candidates.} Symmetric-primitive
VRFs~\cite{ESORICS:BDEKKL22} reduce pseudorandomness to hash/PRP assumptions
at the cost of large (tens of KB) proofs; ring-signature-based PQ
VRFs~\cite{EPRINT:Abraham18} achieve both properties from lattice assumptions.
Every current PQ VRF is substantially slower or larger than ECVRF, and
integration with light-client committee sampling remains unsolved. 
\subsubsection{Cryptography for Privacy and Advanced Functionality}
\label{sec:crypto-priv}

\pparagraph{Zero-knowledge proofs}
A non-interactive zero-knowledge (NIZK) argument for $\mathsf{NP}$-relation
$\mathcal{R}$ is $(\Setup, \Prove, \Verify)$.
 
\begin{property}[NIZK security]
\label{prop:zk}
A NIZK satisfies \textbf{completeness} (honest proofs verify),
\textbf{(knowledge) soundness} (no efficient prover convinces $\Verify$ of
$x \notin \mathcal{L}_{\mathcal{R}}$; the knowledge variant requires an
extractor recovering $w$), and \textbf{zero-knowledge} (a simulator produces
proofs indistinguishable from real ones without $w$).
\end{property}
 
\emph{Quantum impact.} Quantum attacks affect different
properties of different systems:
Groth16~\cite{EC:Groth16}, PLONK~\cite{EPRINT:GabWilCio19}
with KZG, Bulletproofs~\cite{SP:BBBPWM18}, and
Halo/Halo2~\cite{C:BDFG21,EPRINT:SHJKGA23} all derive (knowledge)
soundness from pairing or DLP assumptions, and Shor breaks soundness in each
case, enabling accepting proofs for false statements (e.g., unbacked rollup
state transitions or shielded notes minted from nothing).
STARKs~\cite{EPRINT:BBHR18} rely only on hash collision resistance
and FRI low-degree testing, so their soundness degrades only marginally under
Grover and they are treated as PQ-secure today. Note that
\emph{zero-knowledge} typically survives quantum adversaries given a uniform
random tape, but Fiat--Shamir constructions require re-analysis in the QROM
via measure-and-reprogram or online-extraction techniques.
 
\noindent \emph{Post-quantum candidates.} STARKs / hash-based IOPs~\cite{EPRINT:BBHR18}
offer transparent setup and PQ soundness at the cost of larger
($\sim$50--200\,KB) proofs amortizable via recursion. Lattice-based SNARKs
provide publicly verifiable, preprocessing, recursive
constructions~\cite{C:ACLMT22} and sublinear R1CS
arguments~\cite{C:LyuNguPla22,C:NguSei22}; lattice ZK is also tuned for the
range and equality proofs needed for confidential
stablecoin transfers~\cite{C:ESLL19,PKC:GZGPX25}. MPC-in-the-head systems
(Ligero, Aurora, Picnic)~\cite{CCS:AHIV17,EC:BCRSVW19,CCS:CDGORR17} yield proofs from symmetric primitives at larger
sizes but minimal assumptions.
 
\pparagraph{Commitment schemes}
$(\Com, \Open)$ binds a committer to $m$ without revealing it.
 
\begin{property}[Commitment security]
\label{prop:com}
A commitment is \textbf{hiding} (commitments to $m_0, m_1$ are
indistinguishable) and \textbf{binding} (no efficient $\adv$ opens the same
commitment to two distinct values). Each property may be \emph{computational}
or \emph{statistical}; the two cannot both be statistical.
\end{property}
 
\noindent \emph{Quantum impact.}
\textbf{Pedersen commitments}~\cite{C:Pedersen91} are statistically hiding and
computationally binding under DLP; \emph{binding falls to Shor}, letting an
adversary equivocate a single commitment to multiple values: forging
proof-of-reserves or double-spending confidential transfers in Monero,
MimbleWimble, Zether~\cite{FC:BAZB20}, and bulletproof-based systems. The
statistical hiding property survives. \textbf{Hash-based commitments}
$c = \hashf(m \,\|\, r)$ are computationally hiding and statistically binding
in the random-oracle model; binding degrades to $\approx 2^{\lambda/2}$ under
Grover (acceptable for $\lambda=256$), and hiding remains comfortable when
$|r| \geq 256$ bits. \textbf{KZG/polynomial commitments} are binding under
$q$-SDH/pairing assumptions and broken by Shor, affecting PLONK-style proofs
and Ethereum EIP-4844 blobs.
 
\noindent \emph{Post-quantum candidates.} Hash-based commitments are a natural drop-in
for nullifier and note commitments in Zcash/Aztec-style pools. Lattice-based
SIS commitments~\cite{C:AttLyuSei20} are homomorphic and integrate with
lattice ZK proofs of opened-value arithmetic. Code-based commitments from LPN
exist but are uncommon in production stacks.
 
\pparagraph{Encryption and AEAD}
Symmetric AEAD targets indistinguishability under chosen-ciphertext attack
and integrity of ciphertext. Symmetric ciphers face Grover-style brute-force
key search: AES-128 retains only $\approx 64$-bit quantum security in a
strong query model and AES-256 retains $\approx 128$, motivating 256-bit
keys post-quantum. Under \emph{superposition queries} to a keyed primitive
(an unrealistic model for blockchain settings but consequential for
standardization), Kaplan et al. \cite{C:KLLN16} showed Simon's algorithm breaks
several MAC and mode constructions including CBC-MAC in polynomial time.

\section{Cost Evidence and Benchmarking Details}
\label{app:costs}

Cost figures are meaningful only together with their execution model. Table~\ref{tab:costs} separates deployed precompiles, calldata-only calculations, draft protocol costs, and measured ordinary-EVM verification. The EIP figures are proposals, not deployed Ethereum prices. The \emph{poqeth} values are measured gas costs for specific schemes and NIST-level-1 parameter choices~\cite{ASIACCS:KSKK25}.

\begin{table*}[t]
\centering
\scriptsize
\setlength{\tabcolsep}{3.5pt}
\begin{tabularx}{\textwidth}{@{}p{2.2cm}p{2.15cm}p{2.45cm}p{2.45cm}Y@{}}
\toprule
\textbf{Mechanism} & \textbf{Key/signature} & \textbf{Verification/data cost} & \textbf{Evidence type} & \textbf{Stablecoin interpretation} \\
\midrule
ECDSA/secp256k1 & 65-byte signature & 3,000 gas for \texttt{ecrecover}; about 1,040 calldata gas & Deployed precompile and calldata schedule~\cite{ethereumYellowPaper,eip2028Calldata} & Current account and administrator baseline \\
ML-DSA-44 calldata & 1,312-byte public key; 2,420-byte signature & About 38,720 calldata gas for non-zero signature bytes & Calculation from FIPS 204 and EIP-2028~\cite{nistFips204,eip2028Calldata} & Excludes verification, public-key storage, aggregation, and transaction overhead \\
EIP-8051 & 20,512-byte expanded verifier key; 2,420-byte signature & 4,500 gas proposed & Draft specialized precompile~\cite{eip8051} & Low price depends on protocol support and expanded key representation \\
EIP-8164 & 1,312-byte key; 2,420-byte signature & 50,000 intrinsic gas proposed & Draft native account rule~\cite{eip8164} & Includes an account migration mechanism but not contract, bridge, or validator migration \\
\emph{poqeth} W-OTS+ & Parameter-specific & 222,114 on-chain; 126,095 Naysayer gas & Measured EVM implementation~\cite{ASIACCS:KSKK25} & Optimistic verification lowers cost but adds dispute and trust assumptions \\
\emph{poqeth} XMSS & Parameter-specific & 4,363,623 on-chain; 594,572 Naysayer gas & Measured EVM implementation~\cite{ASIACCS:KSKK25} & Stateful signing can be managed off-chain, but verification remains expensive \\
\emph{poqeth} SPHINCS+ & Parameter-specific & 11,617,690 on-chain; 693,721 Naysayer gas & Measured EVM implementation~\cite{ASIACCS:KSKK25} & Conservative hash assumptions do not imply cheap contract verification \\
\emph{poqeth} MAYO & Parameter-specific & 938,752,492 on-chain; 107,634,064 Naysayer gas & Measured EVM implementation~\cite{ASIACCS:KSKK25} & Compact signatures can still require prohibitive verifier arithmetic \\
SLH-DSA & Parameter-dependent; signatures in the kilobyte-to-tens-of-kilobytes range & Workload and implementation dependent & NIST standard~\cite{nistFips205} & Plausible for cold recovery, generally unsuitable for frequent L1 messages without aggregation \\
\bottomrule
\end{tabularx}
\caption{Representative post-quantum verification and data costs. Values are not directly comparable across execution models.}
\label{tab:costs}
\end{table*}

\pparagraph{Stablecoin benchmark scenarios}
A stablecoin benchmark should evaluate complete operations: mint or burn authorization; threshold custody withdrawal; bridge attestation and redemption; oracle update and retry; user or dormant-balance migration claims; private transfer and selective disclosure; and rollup settlement. Each benchmark should report bytes, gas or native verifier cost, data-availability overhead, signer throughput, end-to-end latency, HSM/MPC support, wallet memory, audit evidence, and behavior under signer or relayer failure.

%% file: abbrev3.bib
@string{ieee =                  {IEEE}}

@string{acm =                   "{ACM}"}


%% file: references.bib
@misc{albarat2024sokpublicsharding,
  title   = {SoK: Public Blockchain Sharding},
  author  = {Al Barat, Md Mohaimin and Li, Shaoyu and Du, Changlai and Hou, Y. Thomas and Lou, Wenjing},
  year    = {2024},
  howpublished = {arXiv preprint arXiv:2405.20521},
  url = {https://arxiv.org/abs/2405.20521},
}

@misc{babbush2026prx,
  author       = {Babbush, Ryan and Zalcman, Adam and Gidney, Craig and Broughton, Michael and Khattar, Tanuj and Neven, Hartmut and Bergamaschi, Thiago and Drake, Justin and Boneh, Dan},
  title        = {Securing Elliptic Curve Cryptocurrencies against Quantum Vulnerabilities: Resource Estimates and Mitigations},
  howpublished = {PRX Quantum, accepted 6 July 2026},
  year         = {2026},
  doi          = {10.1103/j3xf-bw18},
  url          = {https://journals.aps.org/prxquantum/accepted/10.1103/j3xf-bw18}
}

@misc{baseri2025quantumrisk,
  author       = {Baseri, Yaser and Hafid, Abdelhakim and Shahsavari, Yahya and Makrakis, Dimitrios and Khodaiemehr, Hassan},
  title        = {Blockchain Security Risk Assessment in Quantum Era, Migration Strategies and Proactive Defense},
  howpublished = {arXiv:2501.11798},
  year         = {2025},
  doi          = {10.48550/arXiv.2501.11798},
  url          = {https://arxiv.org/abs/2501.11798}
}

@misc{bip360,
  author       = {Beast, Hunter and Heilman, Ethan and Duke, Isabel Foxen},
  title        = {{BIP}-360: Pay-to-Merkle-Root ({P2MR})},
  howpublished = {Bitcoin Improvement Proposals},
  year         = {2024},
  note         = {Draft; assigned 18 December 2024; version updated 29 July 2026},
  url          = {https://github.com/bitcoin/bips/blob/master/bip-0360.mediawiki}
}

@misc{bip361,
  author       = {Lopp, Jameson and Papathanasiou, Christian and Smith, Ian and Ross, Joe and Vaile, Steve and Dallaire-Demers, Pierre-Luc},
  title        = {{BIP}-361: Post Quantum Migration and Legacy Signature Sunset},
  howpublished = {Bitcoin Improvement Proposals},
  year         = {2026},
  note         = {Draft; assigned 11 February 2026},
  url          = {https://github.com/bitcoin/bips/blob/master/bip-0361.mediawiki}
}

@misc{brassard1997quantum,
  title={Quantum algorithm for the collision problem},
  author={Brassard, Gilles and Hoyer, Peter and Tapp, Alain},
  year={1997},
  howpublished = {arXiv preprint arXiv:quant-ph/9705002},
  url = {https://arxiv.org/abs/quant-ph/9705002},
}

@article{buser2023exoticsignatures,
  title     = {A Survey on Exotic Signatures for Post-Quantum Blockchain: Challenges and Research Directions},
  author    = {Buser, Maxime and Dowsley, Rafael and Esgin, Muhammed and Gritti, Clementine and Kasra Kermanshahi, Shabnam and Kuchta, Veronika and Legrow, Jason and Liu, Joseph and Phan, Raphael and Sakzad, Amin},
  journal   = {ACM Computing Surveys},
  volume    = {55},
  number    = {12},
  pages     = {1--32},
  year      = {2023},
  publisher = {ACM}
}

@misc{circleCCTPDeprecation2025,
  author       = {{Circle Internet Group}},
  title        = {{CCTP V1} Deprecation: {CCTP V2} Is Now the Canonical {CCTP}},
  howpublished = {Circle blog},
  year         = {2025},
  month        = nov,
  day          = {14},
  url          = {https://www.circle.com/blog/cctp-version-updates}
}

@misc{circleInco2024,
  author       = {{Circle Research} and {Inco Network}},
  title        = {Unveiling the Confidential {ERC-20} Framework: Compliant Privacy on Public Blockchains Using {FHE}},
  howpublished = {Circle Research blog and whitepaper},
  year         = {2024},
  month        = oct,
  day          = {28},
  url          = {https://www.circle.com/blog/confidential-erc-20-framework-for-compliant-on-chain-privacy}
}

@misc{circlePQRoadmap2026,
  author       = {Belenkiy, Mira and Le, Duc V. and Liao, Gordon and Sehrawat, Vipin Singh and Rotaru, Dragos and Gorbunov, Sergey and Sheth, Milap and Logelin, Jay and De Abreu, Anthony and Boneh, Dan},
  title        = {Circle's Post-Quantum Security Roadmap: Securing Blockchains, Smart Contracts, and Digital Assets for the Quantum Era},
  howpublished = {Circle technical roadmap},
  year         = {2026},
  month        = may,
  url          = {https://6778953.fs1.hubspotusercontent-na1.net/hubfs/6778953/PDFs/quantum_paper.pdf}
}

@misc{circleStablecoinEVM,
  author       = {{Circle Internet Financial}},
  title        = {Stablecoin Smart Contracts on {EVM}-Compatible Blockchains},
  howpublished = {GitHub repository},
  year         = {2026},
  note         = {Accessed 6 August 2026},
  url          = {https://github.com/circlefin/stablecoin-evm}
}

@misc{circleXReserve2025,
  author       = {{Circle Internet Group}},
  title        = {Introducing Circle xReserve: Enabling Interoperable {USDC}-Backed Stablecoins for Blockchains},
  howpublished = {Circle blog},
  year         = {2025},
  month        = nov,
  day          = {18},
  url          = {https://www.circle.com/blog/introducing-circle-xreserve}
}

@misc{eip2028Calldata,
	title        = {EIP-2028: Transaction data gas cost reduction},
	author       = {Buterin, Vitalik and Conner, Eric and Dudley, John and Slipper, Matthew and Norden, Ian and Bakhta, Abdelhamid},
	year         = {2019},
	howpublished = {Ethereum Improvement Proposals},
	url          = {https://eips.ethereum.org/EIPS/eip-2028}
}

@misc{eip8051,
  author       = {Dubois, Renaud and Masson, Simon},
  title        = {{EIP}-8051: Precompile for {ML-DSA} Signature Verification},
  howpublished = {Ethereum Improvement Proposals},
  year         = {2025},
  note         = {Draft; created 15 October 2025},
  url          = {https://eips.ethereum.org/EIPS/eip-8051}
}

@misc{eip8164,
  author       = {Markou, Gregory and Prestwich, James},
  title        = {{EIP}-8164: Native Key Delegation for {EOAs}},
  howpublished = {Ethereum Improvement Proposals},
  year         = {2026},
  note         = {Draft; created 17 February 2026},
  url          = {https://eips.ethereum.org/EIPS/eip-8164}
}

@misc{ethereumPQ2026,
  author       = {{Ethereum Foundation}},
  title        = {Post-Quantum Ethereum},
  howpublished = {Ethereum research roadmap},
  year         = {2026},
  note         = {Accessed 6 August 2026},
  url          = {https://pq.ethereum.org/}
}

@misc{ethereumYellowPaper,
	title        = {Ethereum: A Secure Decentralised Generalised Transaction Ledger},
	author       = {Wood, Gavin},
	year         = {2024},
	howpublished = {Ethereum Yellow Paper},
	url          = {https://ethereum.github.io/yellowpaper/paper.pdf},
	note         = {Used for Ethereum precompile and gas-cost background}
}

@article{fernandezcarames2020pqblockchain,
  title     = {Towards Post-Quantum Blockchain: A Review on Blockchain Cryptography Resistant to Quantum Computing Attacks},
  author    = {Fernandez-Carames, Tiago M. and Fraga-Lamas, Paula},
  journal   = {IEEE Access},
  volume    = {8},
  pages     = {21091--21116},
  year      = {2020},
  publisher = {IEEE}
}

@misc{geniusActText2025,
  author       = {{United States Congress}},
  title        = {Text of {S. 1582}: {GENIUS Act}},
  howpublished = {119th Congress, Public Law 119-27},
  year         = {2025},
  url          = {https://www.congress.gov/bill/119th-congress/senate-bill/1582/text}
}

@misc{ghosh2025quantumblockchain,
  title   = {Quantum Blockchain Survey: Foundations, Trends, and Gaps},
  author  = {Ghosh, Saurav and Mishu, Niloy Deb Roy},
  year    = {2025},
  howpublished = {arXiv preprint arXiv:2507.13720},
  url = {https://arxiv.org/abs/2507.13720},
}

@article{joseph2022pqmigrationIntro,
  title     = {Transitioning organizations to post-quantum cryptography},
  author    = {Joseph, David and Misoczki, Rafael and Manzano, Marc and Tricot, Joe and Pinuaga, Fernando Dominguez and Lacombe, Olivier and Leichenauer, Stefan and Hidary, Jack and Venables, Phil and Hansen, Royal},
  journal   = {Nature},
  volume    = {605},
  number    = {7909},
  pages     = {237--243},
  year      = {2022},
  publisher = {Nature Publishing Group}
}

@misc{li2026stablecoinretail,
  author       = {Li, Yuquan and Xiang, Yuexin and Wang, Qin and Yuen, Tsz Hon and Deppeler, Andreas and Yu, Jiangshan},
  title        = {SoK: Stablecoins in Retail Payments},
  howpublished = {arXiv:2601.00196},
  year         = {2026},
  doi          = {10.48550/arXiv.2601.00196},
  url          = {https://arxiv.org/abs/2601.00196}
}

@misc{ling2025stablecoinlego,
  author       = {Ling, Shengchen and Du, Yuefeng and Zhou, Yajin and Wu, Lei and Wang, Cong and Jia, Xiaohua and Yan, Houmin},
  title        = {SoK: Stablecoin Designs, Risks, and the Stablecoin LEGO},
  howpublished = {arXiv:2506.17622},
  year         = {2025},
  doi          = {10.48550/arXiv.2506.17622},
  url          = {https://arxiv.org/abs/2506.17622}
}

@misc{mallick2025quantumdisruption,
  author       = {Mallick, Tushin and Zeldin, Maya and Cenk, Murat and Nita-Rotaru, Cristina},
  title        = {Quantum Disruption: An SoK of How Post-Quantum Attackers Reshape Blockchain Security and Performance},
  howpublished = {arXiv:2512.13333},
  year         = {2025},
  doi          = {10.48550/arXiv.2512.13333},
  url          = {https://arxiv.org/abs/2512.13333}
}

@misc{mica2023,
  author       = {{European Parliament and Council of the European Union}},
  title        = {Regulation ({EU}) 2023/1114 on Markets in Crypto-Assets},
  howpublished = {Official Journal of the European Union},
  year         = {2023},
  url          = {https://eur-lex.europa.eu/eli/reg/2023/1114}
}

@misc{mosca2018cybersecurity,
	title        = {Cybersecurity in an Era with Quantum Computers: Will We Be Ready?},
	author       = {Mosca, Michele},
	year         = {2018},
	howpublished = {Cryptology ePrint Archive, Paper 2015/1075},
	url          = {https://eprint.iacr.org/2015/1075}
}

@misc{nist2023nccoequantumreadiness,
  title        = {Migration to Post-Quantum Cryptography: Quantum Readiness: Cryptographic Discovery},
  author       = {{National Cybersecurity Center of Excellence} and {National Institute of Standards and Technology}},
  year         = {2023},
  howpublished = {NIST SP 1800-38B Preliminary Draft},
  url          = {https://www.nccoe.nist.gov/projects/migration-post-quantum-cryptography}
}

@misc{nistAdditionalSigs2026,
  author       = {{National Institute of Standards and Technology}},
  title        = {Nine Candidates Advance to the Third Round of the Additional Digital Signatures for the {PQC} Standardization Process},
  howpublished = {NIST News},
  year         = {2026},
  month        = may,
  day          = {14},
  note         = {NIST Internal Report 8610},
  url          = {https://www.nist.gov/news-events/news/2026/05/nine-candidates-advance-third-round-additional-digital-signatures-pqc}
}

@techreport{nistCryptoAgility2026,
  author      = {Barker, Elaine and Chen, Lily and Cooper, David and Moody, Dustin and Regenscheid, Andrew and Souppaya, Murugiah and Newhouse, William and Housley, Russ and Turner, Sean and Barker, William and Kent, Karen},
  title       = {Considerations for Achieving Crypto Agility: Strategies and Practices},
  institution = {National Institute of Standards and Technology},
  type        = {NIST Cybersecurity White Paper 39 Update 1},
  year        = {2026},
  month       = jun,
  doi         = {10.6028/NIST.CSWP.39-upd1},
  url         = {https://csrc.nist.gov/pubs/cswp/39/upd1/considerations-for-achieving-crypto-agility/final}
}

@misc{nistFips203,
  author       = {{National Institute of Standards and Technology}},
  title        = {{FIPS 203}: Module-Lattice-Based Key-Encapsulation Mechanism Standard},
  howpublished = {Federal Information Processing Standards Publication 203},
  year         = {2024},
  doi          = {10.6028/NIST.FIPS.203},
  url          = {https://csrc.nist.gov/pubs/fips/203/final}
}

@misc{nistFips204,
	title        = {FIPS 204: Module-Lattice-Based Digital Signature Standard},
	author       = {{National Institute of Standards and Technology}},
	year         = {2024},
	howpublished = {Federal Information Processing Standards Publication 204},
	url          = {https://csrc.nist.gov/pubs/fips/204/final}
}

@misc{nistFips205,
	title        = {FIPS 205: Stateless Hash-Based Digital Signature Standard},
	author       = {{National Institute of Standards and Technology}},
	year         = {2024},
	howpublished = {Federal Information Processing Standards Publication 205},
	url          = {https://csrc.nist.gov/pubs/fips/205/final}
}

@misc{nistHQC2025,
  author       = {{National Institute of Standards and Technology}},
  title        = {{HQC} Announced as a Fourth-Round Selection in the {NIST PQC} Standardization Process},
  howpublished = {NIST News},
  year         = {2025},
  month        = mar,
  day          = {11},
  note         = {See NIST Internal Report 8545},
  url          = {https://www.nist.gov/news-events/news/2025/03/nist-pqc-standardization-process-hqc-announced-4th-round-selection}
}

@misc{nistPQCProject2026,
  author = {{National Institute of Standards and Technology}},
  title = {Post-Quantum Cryptography Project},
  year = {2026},
  howpublished = {NIST Computer Security Resource Center},
  url = {https://csrc.nist.gov/Projects/Post-Quantum-Cryptography},
  note = {Accessed August 2026}
}

@techreport{nistTransition2024,
  author      = {Moody, Dustin and Perlner, Ray and Regenscheid, Andrew and Robinson, Angela and Cooper, David},
  title       = {Transition to Post-Quantum Cryptography Standards},
  institution = {National Institute of Standards and Technology},
  type        = {NIST Internal Report 8547, Initial Public Draft},
  year        = {2024},
  doi         = {10.6028/NIST.IR.8547.ipd},
  url         = {https://csrc.nist.gov/pubs/ir/8547/ipd}
}

@misc{notland2024sokbridging,
  author       = {Notland, Jakob Svennevik and Li, Jinguye and Nowostawski, Mariusz and Haro, Peter Halland},
  title        = {SoK: Cross-Chain Bridging Architectural Design Flaws and Mitigations},
  howpublished = {arXiv:2403.00405},
  year         = {2024},
  doi          = {10.48550/arXiv.2403.00405},
  url          = {https://arxiv.org/abs/2403.00405}
}

@misc{occGenius2026,
  author       = {{Office of the Comptroller of the Currency}},
  title        = {{GENIUS Act} Regulations: Notice of Proposed Rulemaking},
  howpublished = {OCC Bulletin 2026-3},
  year         = {2026},
  month        = feb,
  day          = {25},
  url          = {https://www.occ.treas.gov/news-issuances/bulletins/2026/bulletin-2026-3.html}
}

@misc{occGeniusAml2026,
  author       = {{Office of the Comptroller of the Currency}},
  title        = {{GENIUS Act}: Anti-Money Laundering/Countering the Financing of Terrorism and Sanctions Compliance: Notice of Proposed Rulemaking},
  howpublished = {OCC Bulletin 2026-28},
  year         = {2026},
  month        = jun,
  day          = {22},
  url          = {https://www.occ.treas.gov/news-issuances/bulletins/2026/bulletin-2026-28.html}
}

@misc{qi2024bitcoinl2,
  title   = {SoK: Bitcoin Layer Two (L2)},
  author  = {Qi, Minfeng and Wang, Qin and Wang, Zhipeng and Schneider, Manvir and Zhu, Tianqing and Chen, Shiping and Knottenbelt, William and Hardjono, Thomas},
  year    = {2024},
  howpublished = {arXiv preprint arXiv:2409.02650},
  url = {https://arxiv.org/abs/2409.02650},
}

@article{simon1997power,
	title     = {On the Power of Quantum Computation},
	author    = {Simon, Daniel R.},
	journal   = {SIAM Journal on Computing},
	volume    = {26},
	number    = {5},
	pages     = {1474--1483},
	year      = {1997},
	publisher = {SIAM}
}

@misc{yang2024pqquantumblockchains,
  author       = {Yang, Zebo and Alfauri, Haneen and Farkiani, Behrooz and Jain, Raj and Di Pietro, Roberto and Erbad, Aiman},
  title        = {A Survey and Comparison of Post-Quantum and Quantum Blockchains},
  howpublished = {arXiv:2409.01358},
  year         = {2024},
  doi          = {10.48550/arXiv.2409.01358},
  url          = {https://arxiv.org/abs/2409.01358}
}

@misc{zcashzip2005,
  title        = {ZIP 2005: Orchard Quantum Recoverability},
  author       = {Hopwood, Daira-Emma and Grigg, Jack},
  year         = {2025},
  month        = mar,
  day          = {31},
  howpublished = {Zcash Improvement Proposals},
  url          = {https://zips.z.cash/zip-2005},
  note         = {Status: Proposed; Category: Consensus}
}

@misc{zhang2022sokdecentralization,
  title   = {SoK: Blockchain Decentralization},
  author  = {Zhang, Luyao and Ma, Xinshi and Liu, Yulin},
  year    = {2022},
  howpublished = {arXiv preprint arXiv:2205.04256},
  url = {https://arxiv.org/abs/2205.04256},
}

@misc{zhang2025stablecoinsrwa,
  author       = {Zhang, Luyao},
  title        = {SoK: Stablecoins for Digital Transformation -- Design, Metrics, and Application with Real World Asset Tokenization as a Case Study},
  howpublished = {arXiv:2508.02403},
  year         = {2025},
  doi          = {10.48550/arXiv.2508.02403},
  url          = {https://arxiv.org/abs/2508.02403}
}

@misc{luo2026pointadd,
  author       = {Luo, Han and Yang, Ziyi and Luo, Jingquan and Wang, Ziruo and Su, Yuexin and Sun, Xiaoming and Li, Lvzhou and Li, Tongyang},
  title        = {Quantum Algorithm for Elliptic Curve Discrete Logarithms with Space-Efficient Point Addition},
  howpublished = {arXiv preprint arXiv:2607.13816},
  year         = {2026},
  month        = jul,
  doi          = {10.48550/arXiv.2607.13816},
  url          = {https://arxiv.org/abs/2607.13816}
}

@techreport{nistSP800208,
  author      = {Cooper, David and Apon, Daniel and Dang, Quynh and Davidson, Michael and Dworkin, Morris and Miller, Carl},
  title       = {Recommendation for Stateful Hash-Based Signature Schemes},
  institution = {National Institute of Standards and Technology},
  type        = {NIST Special Publication},
  number      = {800-208},
  year        = {2020},
  month       = oct,
  doi         = {10.6028/NIST.SP.800-208},
  url         = {https://csrc.nist.gov/pubs/sp/800/208/final}
}

@techreport{ietfTLSHybridMLKEM2026,
  author      = {Kwiatkowski, Kris and Kampanakis, Panos and Westerbaan, Bas and Stebila, Douglas},
  title       = {Post-quantum hybrid {ECDHE-MLKEM} Key Agreement for {TLSv1.3}},
  institution = {Internet Engineering Task Force},
  type        = {Internet-Draft},
  number      = {draft-ietf-tls-ecdhe-mlkem-05},
  year        = {2026},
  month       = may,
  note        = {Work in Progress, published 26 May 2026},
  url         = {https://datatracker.ietf.org/doc/draft-ietf-tls-ecdhe-mlkem/}
}

@misc{algorandPQRoadmap2026,
  author       = {{Algorand Foundation}},
  title        = {Algorand Post-Quantum Cryptography Roadmap},
  howpublished = {Algorand Foundation technical blog},
  year         = {2026},
  month        = jun,
  day          = {18},
  url          = {https://algorand.co/blog/algorand-post-quantum-cryptography-roadmap}
}

@misc{starknetPQRoadmap2026,
  author       = {{StarkWare}},
  title        = {The Architecture Advantage: {Starknet}'s Post-Quantum Roadmap},
  howpublished = {StarkWare blog},
  year         = {2026},
  month        = jun,
  day          = {30},
  url          = {https://starkware.co/blog/the-architecture-advantage-starknets-quantum-readiness-roadmap/}
}

@misc{suiQuantum2025,
  author       = {{Mysten Labs}},
  title        = {Securing {Sui} in the Quantum Computing Era},
  howpublished = {Sui blog},
  year         = {2025},
  month        = apr,
  day          = {10},
  url          = {https://blog.sui.io/post-quantum-computing-cryptography-security/}
}

@misc{solanaQuantum2026,
  author       = {{Solana Foundation}},
  title        = {{Solana}'s Quantum Readiness},
  howpublished = {Solana Foundation},
  year         = {2026},
  month        = apr,
  day          = {27},
  url          = {https://solana.com/news/quantum-readiness}
}

@misc{coinbaseQuantumCouncil2026,
  author       = {{Coinbase}},
  title        = {Coinbase Quantum Advisory Council Publishes Position Paper on Quantum Computing and Blockchain},
  howpublished = {Coinbase blog},
  year         = {2026},
  month        = apr,
  day          = {21},
  url          = {https://www.coinbase.com/blog/coinbase-quantum-advisory-council-publishes-position-paper-on-quantum-computing-and-blockchain}
}

@misc{coinbaseAbandonedCoins2026,
  author       = {Lindell, Yehuda},
  title        = {Coinbase Quantum Advisory Council: Post-Quantum Migration and Abandoned Coins},
  howpublished = {Coinbase blog},
  year         = {2026},
  month        = jun,
  day          = {11},
  url          = {https://www.coinbase.com/blog/coinbase-quantum-advisory-council-post-quantum-migration-and-abandoned-coins}
}

@misc{blockstreamLiquidPQ2026,
  author       = {{Blockstream Team}},
  title        = {Blockstream Research Demonstrates Quantum-Resistant Transaction Signing on {Liquid} Using {Simplicity} Smart Contracts},
  howpublished = {Blockstream Research blog},
  year         = {2026},
  month        = mar,
  day          = {3},
  url          = {https://blog.blockstream.com/blockstream-research-demonstrates-quantum-resistant-transaction-signing-on-liquid-using-simplicity-smart-contracts/}
}
